\documentclass[12pt]{article}
\usepackage[T1]{fontenc}
\usepackage{geometry}                % See geometry.pdf to learn the layout options. There are lots.
\usepackage{graphicx}
\usepackage{subcaption}
\usepackage{amssymb}
\usepackage{bm}
\usepackage{amsmath}
\usepackage{epstopdf}
\usepackage{color}
\usepackage{verbatim}
\usepackage{hyperref}
\usepackage{url}
\usepackage{soul}
\begin{document}
\title{Study of lockdown/testing mitigation strategies on stochastic SIR model and its comparison with South Korea, Germany and New York data}
\author{Priyanka$^\dagger$ and Vicky Verma$^\ddagger$}
\date{%
    $^\dagger$Department of Physics and Center for Soft Matter and Biological Physics, Virginia Tech, Blacksburg, VA 24061-0435, USA \\%
    $^\ddagger$Department of Mechanical and Aerospace Engineering, University of California, San Diego, 92093-0411, USA\\[2ex]%
    \today
}
\maketitle
\begin{abstract}

We are currently facing a highly critical case of a world-wide pandemic. The novel coronavirus (SARS-CoV-2, a.k.a. COVID-19) has proved to be extremely contagious and the original outbreak from Asia has now spread to all continents. This situation will fruitfully profit from the study in regards of the spread of the virus, assessing effective countermeasures to weight the impact of the adopted strategies. The standard Susceptible-Infectious-Recovered (SIR) model is a very successful and widely used mathematical model for predicting the spread of an epidemic. We adopt the SIR model on a random network and extend the model to include control strategies {\em lockdown} and {\em testing} -- two often employed mitigation strategies. The ability of these strategies in controlling the pandemic spread is investigated by varying the effectiveness with which they are implemented. The possibility of a second outbreak is evaluated in detail after the mitigation strategies are withdrawn. We notice that, in any case, a sudden interruption of such mitigation strategies will likely induce a resurgence of a second outbreak, whose peak will be correlated to the number of susceptible individuals. In fact, we find that a population will remain vulnerable to the infection until the herd immunity is achieved. We also test our model with real statistics and information on the epidemic spread in South Korea, Germany, and New York and find a remarkable agreement with the simulation data.

%  In the current scenario, there is a need for a detailed understanding of the effect of control measures taken during the outbreak of COVID-19. The most accepted measure is lockdown; hence in this article, we have shown the impact of mitigation on the growth of infections in various situations. Our finding suggests that a population is always vulnerable to infection until it achieves herd immunity. Our studies show the immediate end of any mitigation strategies resurgence the second outbreak with peak value determined by the number of susceptibles. We have observed that lockdown alone is never a sufficient strategy to contain the infections, and testing is a much better approach. Furthermore, we collect South Korea, Germany, and New York data to reproduce the mitigated growth of infections using a modified stochastic three state compartmental model on a random Network.
\end{abstract}

%\linenumbers
\section{Introduction}
In the recent outbreak of the highly contagious novel coronavirus (SARS-CoV-2 a.k.a COVID-19), which emerged out of the city of Wuhan, the infection has spread mainly through people-to-people contacts. Consequently, the policies have relied on limiting the contacts by closing down schools, banning mass gatherings, and encouraging individuals to maintain sufficient distance from others while in the public spaces, and mandating mask-wearing. These mitigation measures help to slowdown the spread, but to achieve sufficient control over it {\em lockdowns} have been imposed by many countries, whereby people are restricted to their homes and the outside movements are restricted to meet only the most necessary needs. The ideal scenario of a complete lockdown is practically impossible to achieve, as people go out to buy food and medical supplies. As a result, there is a finite probability with which people follow the lockdown, and the value of one represents the ideal case of complete lockdown.    

Another controlling strategy is {\em testing}, where the infection can be identified and isolated through testing. The testing measure has been employed effectively by South Korea to contain the spread of coronavirus without imposing severe social distancing measures used by other countries. The interventions such as the closure of schools and workplaces, banning mass gatherings, mandated mask-wearing can further aide the testing measure. 

The uncontrolled spread of coronavirus can lead to a large portion of the population becoming sick over a span of few days, many of whom will need medical attention at a hospital, and some might succumb to the decease without proper medical care. It is a fact that most hospitals would not be able to handle such large numbers of patients, as they can normally provide care to only a few patients per ten thousand of the population. Therefore, to curb the burden on the health care system and to reduce the mortality rate, making it possible for each individual to receive the needed medical assistance~\cite{Ferguson:2020, Bootsma:2007, Hatchett:2007, Maharaj:2012}, mitigating interventions are required. The lockdown and testing strategies are successful at controlling the spread, and knowing how infection spreads in the system when these strategies are employed would allow us to make proper usage of our resources and to plan to meet the future needs better. Equally important is to know what happens when the controls are lifted, allowing the population to function normally. The lockdown and testing strategies essentially operate by reducing the effective reproduction number, ${\cal{R}}_e$, and a value smaller than one indicates a decline in the number of cases~\cite{Ferguson:2006, Kermack:1997, Brauer:2017}. However, the actual reproduction number ${\cal R}_0$ of the uncontrolled system remains the same. Hence, any decrease in effective reproduction number does not provide the assurance of eliminating the epidemic in this well connected world. Many studies have come up since February 2020 on predicting the growth of the epidemic using mathematical modeling~\cite{Prem:2020, Singh:2020}, as well as numerical simulations with particles and networks \cite{Ferguson:2020}. Further, extensive studies on the effect of social distancing interventions have also been done ~\cite{ Brethouwer:2020, Maharaj:2012, Prem:2020, Ruslan:2020}. Nevertheless, incorporating the various other mitigation strategies in the predictive models in a realistic manner is still not well understood, and further studies are required. 

For such studies, it is well known that stochastic modeling of the epidemic outbreak can be performed using a network of contacts of individuals. A network provides an intuitive way to model the spread of an epidemic described by a graph, where each vertex represents an individual, and the infection spreads from one individual to another through contacts represented by edges \cite{Easley:2010, Brauer:2017}. Designing a suitable network is an essential aspect of the problem and can have a significant impact on the prediction. A variety of network designs with distinct features are available. Among them, random and small-world networks are often used in epidemiological modeling. A small-world network is capable of modeling long- and short-range connections and displays a significant effect on the spread of infection to reach all parts of the world quickly~\cite{Watts:1998, Newman:2002, Keeling:2005}. Some studies concerning COVID-19 show the containment in the spread of the disease by weakening or disconnecting these long-range links \cite{Brethouwer:2020, Ruslan:2020}. A more local approach can be employed for modeling regions that employ travel restrictions, severing large-range connections, and ban large gatherings of people, limiting the number of local contacts per-person. Hence, we believe that for understanding the dynamics of an epidemic in such mitigated conditions, the Erd{\"o}s-R{\'e}nyi random network is an appropriate choice \cite{Erdos:1959}. In this network, each vertex is connected randomly with other vertices with a constant probability, and the probability distribution of the edges is Poissonian~\cite{Erdos:1959}.

The main goal of this study is to predict the growth of the epidemic with the lockdown and the testing strategies and investigate their impact on the second outbreak through numerical simulations. Although the system is assumed to be closed, the influence of the outside world on the final state of the system is also considered by adding external infections. We have shown detailed numerical simulations on a million population to mimic the growth of COVID-19 in South Korea, Germany and New York. These studies on real data uses the approach which includes both lockdown and testing measures in the simulations. Interestingly and importantly, we have performed the numerical simulation by tuning only two-parameters (reflecting the extent of lockdown and the effectiveness of testing) to achieve the desire results. Such studies are helpful to identify effective countermeasures and to weight the impact of the adopted strategies and also make us understand the limitations of the modeling.

The paper is organized as follows. The modeling approach and the comparison with the three state compartmental model for an uncontrolled growth is presented in section \ref{Uncontrol}. Sec.\ref{Sec:Results} contains the epidemic modeling with different mitigation strategies and, further, the impacts of the lockdown and the testing strategies on the epidemic dynamics are investigated in subsections \ref{Sec:Testing} and \ref{Sec:Lockdown}, respectively. In Sec.\ref{Sec:Realdata}, our modeling of the epidemic growth is compared with statistics of the epidemic diffusion in South Korea, Germany and New York. The results are finally discussed in Sec.~\ref{Sec:Concl}.

\section{Susceptible-Infectious-Recovered (SIR) model}
\label{Uncontrol}
The susceptible-infectious-recovered (SIR) model is one of the widely used standard mathematical models for predicting the epidemic growth. The model is suitable for those infections in which recovered individuals develop a long term immunity and are unlikely to get infected the second time, as the infection spreads. In this model, the population is compartmentalized into the groups of susceptible (S), infectious (I), and recovered (R), and their evolution is given by \cite{Kermack:1997, Brauer:2017}
\begin{align} \label{eq:SIR}
 \frac{dS}{dt} &= -\alpha\frac{IS}{N}, \notag \\ 
 \frac{dI}{dt} &=  \alpha\frac{IS}{N} - \gamma I, \notag \\
 \frac{dR}{dt} &= \gamma I,
\end{align}
where the constants $\alpha$ and $\gamma$ are related to the transmission of the infection and the recovery of the infected, respectively, in a population of size $S+I+R = N$. In the above system, each individual interacts with all the other members of the population with small but equal probability, and the infection spreads through the contacts at an average rate of $\tau$. Thus, for a system (e.g. a random network) with a finite number of $\hat{n}$ mean contacts per unit time, we get $\alpha = \tau \hat{n}$.  

The model exhibits a peak in the evolution of the infectious, occurring when $dI/dt = 0$ and ${\cal{R}}_0 \equiv \alpha/\gamma = 1/\tilde{S}$, where $\tilde{S} = S/N$. Subsequently, $dI/dt$ becomes negative, and the number of infectious continues to decrease. The spread stops after $dI/dt$ becomes zero. By this time, every individual in the population may not become infected; this situation is often described as the {\em herd immunity}. The total number of cases for reaching the herd immunity is dependent on ${\cal{R}}_0$ -- the higher the value of ${\cal{R}}_0$, the higher is the number of total infected in the population before the herd immunity is attained.

The standard SIR model represents an ideal system where each individual interacts with all other individuals, a so-called well mixed system. In a real system, however, an individual can have only a finite number of contacts. The interaction between any two individuals can also depend on many other factors such as distance, age group, and profession \cite{Ferguson:2020, Halloran:2008}, etc. Including these details in the SIR model is not straight forward; however, many of the issues can be addressed easily through network modeling, which allows flexibility in manipulating the contacts. Exploiting this flexibility, we aim to implement SIR dynamics on a network and enable the inclusion of lockdown and testing control strategies. 
%As we will show, both of these strategies can be implemented easily on a network model by manipulating the transmission through the edges.% Moreover, in case of COVID-19 SAIR play an important role when it come to controlling strategies, especially testing. 
\begin{figure}[!ht]
   \includegraphics[width=\textwidth]{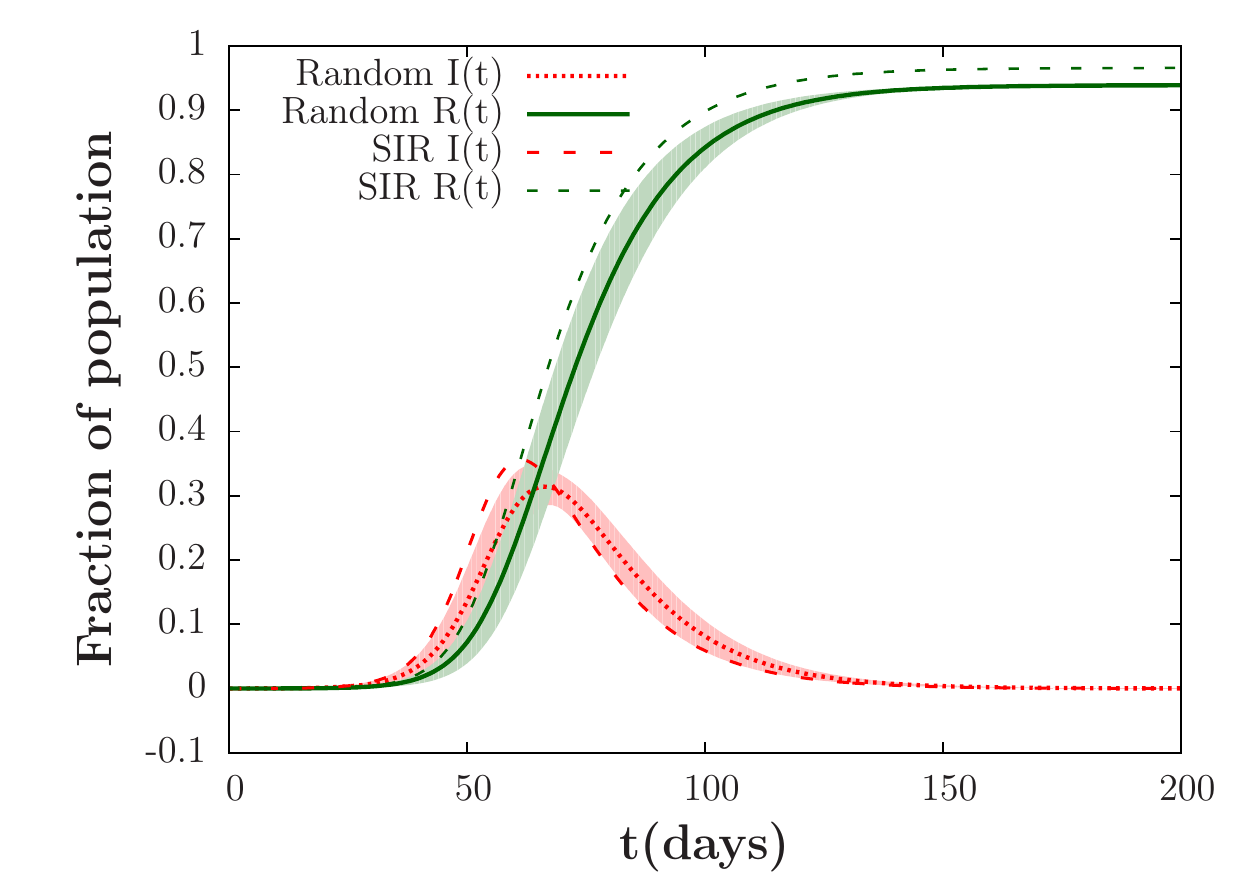}
    \caption{The time evolution of I and R predicted from a network simulation of SIR dynamics as well as that obtained by integrating the continuous SIR model. The curves for random network are averaged over 100 different realizations. The parameters used are $\gamma=1/14$, $\alpha=0.25$ with $I(0)=5$ for a population of size 0.1 million.}
    \label{fig:network_unmitigated}
\end{figure}

For simulating the SIR dynamics on a network, one needs to set the parameters $\alpha$ and $\gamma$ appropriately. These parameters are selected utilizing the available data for the spread of COVID-19. Considering the mean recovery period of approximately $14$ days, $\gamma = 1/14 \, \rm{day^{-1}}$ is assigned. The value of $\alpha$ is calculated from $R_0$ reported in the literature. For COVID-19, ${\cal{R}}_0 = (5.5 - 3.26)$. In many of the results concerning mitigation strategies presented here, ${\cal R}_0 = 3.53$ is chosen, and the corresponding value of $\alpha = 0.25\,\rm{day^{-1}}$. The network is marched in time following Gillespie algorithm~\cite{Vestergaard:2015}. 

In Fig.~\ref{fig:network_unmitigated}, we compare the results obtained for the $10^5$ populations from the continuous SIR model and those from the random network model with $20$ mean connection for the same parameters. The SIR model shows a steep increase in the number of infectious, peaking at $35\%$ of the total population, around $75^{th}$ days after the exposer. Thereafter, the number of infectious decreases and drops to zero beyond $175$ days and the population attains herd immunity with $98\%$ becoming infected. As noted earlier, the percentage of the total infected (recovered) at the end depends strongly on ${\cal{R}}_0$, which corresponds to the rate of the infection spread, $\alpha$. The results from the random network model is qualitatively similar, but some differences can be noticed. First, the infectious curve reaches the peak value a few days later with slightly lower number of infected, at $30 \%$ of the population. Second, the heard immunity is reached after $95\%$ of the population becomes infected, which is slightly lower compared to the SIR model. Overall, the predictions from the two models are in good agreement. 

Using the same parameters, in the section below, we explore the effect of mitigation measures on the spread of the pandemic and try to explore the best strategy for its control.  
 
\section{Results}
\label{Sec:Results}
In this section, we discuss how the effect of the lockdown and the testing can be including in the SIR random network model and investigate the epidemic dynamics in the mitigated system.
 
\subsection{Lockdown}
\label{Sec:Lockdown}

A simple and effective mitigation strategy is to impose lockdown. This approach relies on limiting the contacts between any two individuals, which ultimately leads to a reduction the rate of infection growth. In the ideal case of complete lockdown, all contacts are disrupted, resulting a decline in the number of the infectious as they recover. In a network, the effect of lockdown can be modeled by isolating the nodes. Effectively, an infectious node is also isolated by making the probability of infection transmission to zero through all edges connected to the node. In a realistic system implementing lockdown can also involve many other complexities. For example, all individuals do not go into lockdown immediately; there is a time window over which this happens. Further, all individuals do not remain isolated at the same, and there is a fraction that remains mobile and can still spread the infection. There is another issue of close contacts, such as those between the members of a family living together. These contacts never break, and the presence of an infectious can affect the other members as well. These details can have significant impact on the outcome, but for simplicity we ignore such complications and study the situation in which contacts are immediately cut-off and a randomly picked individuals become isolated. However, we have considered the scenario in which a fraction of the population is mobile even after the lockdown is in place due random picking update in simulation.

\begin{figure}[!ht]
\centering
    \begin{subfigure}[b]{0.5\textwidth}            
     \includegraphics[width=\textwidth]{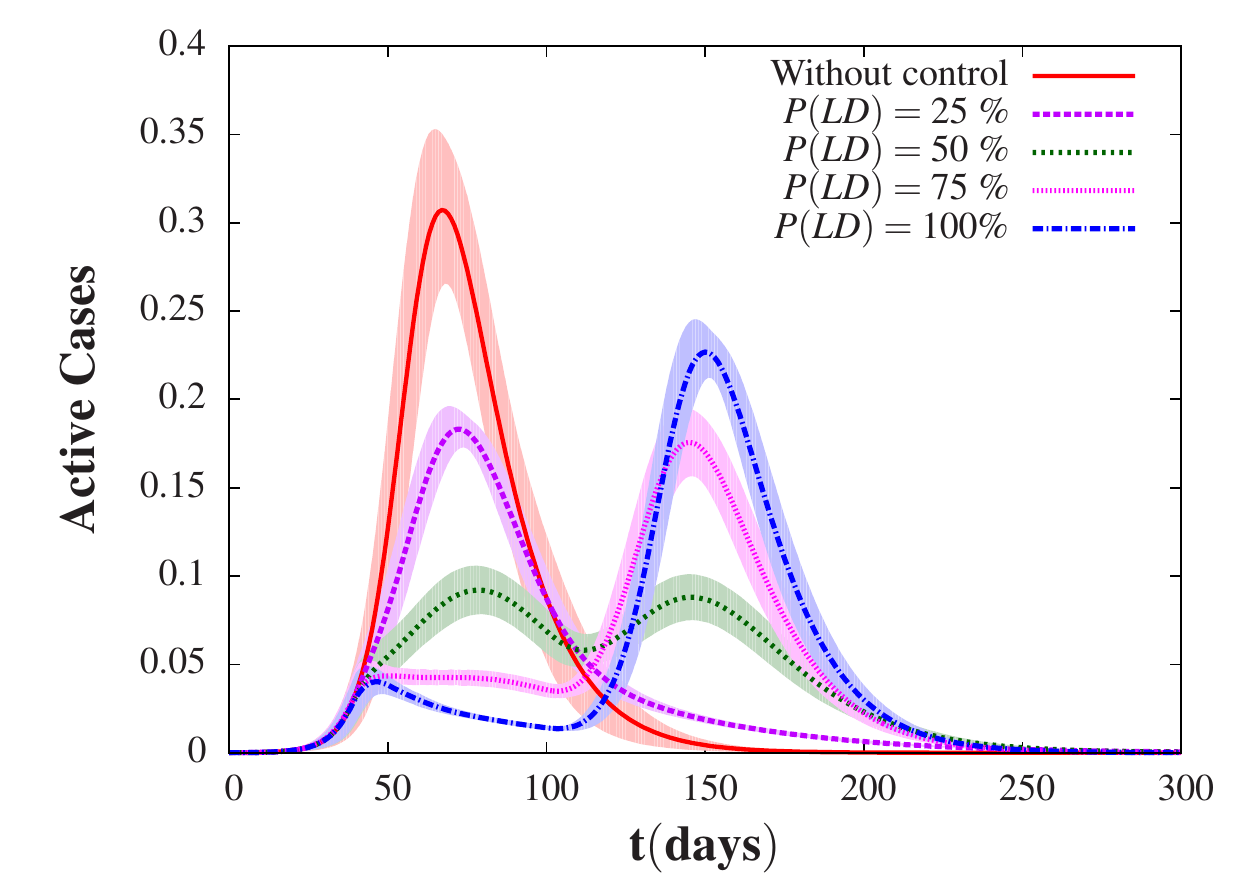}

            \caption{Lockdown started at $I(t) = 0.05$}
            \label{fig:lockdown_a}
    \end{subfigure}%
     %add desired spacing between images, e. g. ~, \quad, \qquad etc.
    \begin{subfigure}[b]{0.5\textwidth}
         \includegraphics[width=\textwidth]{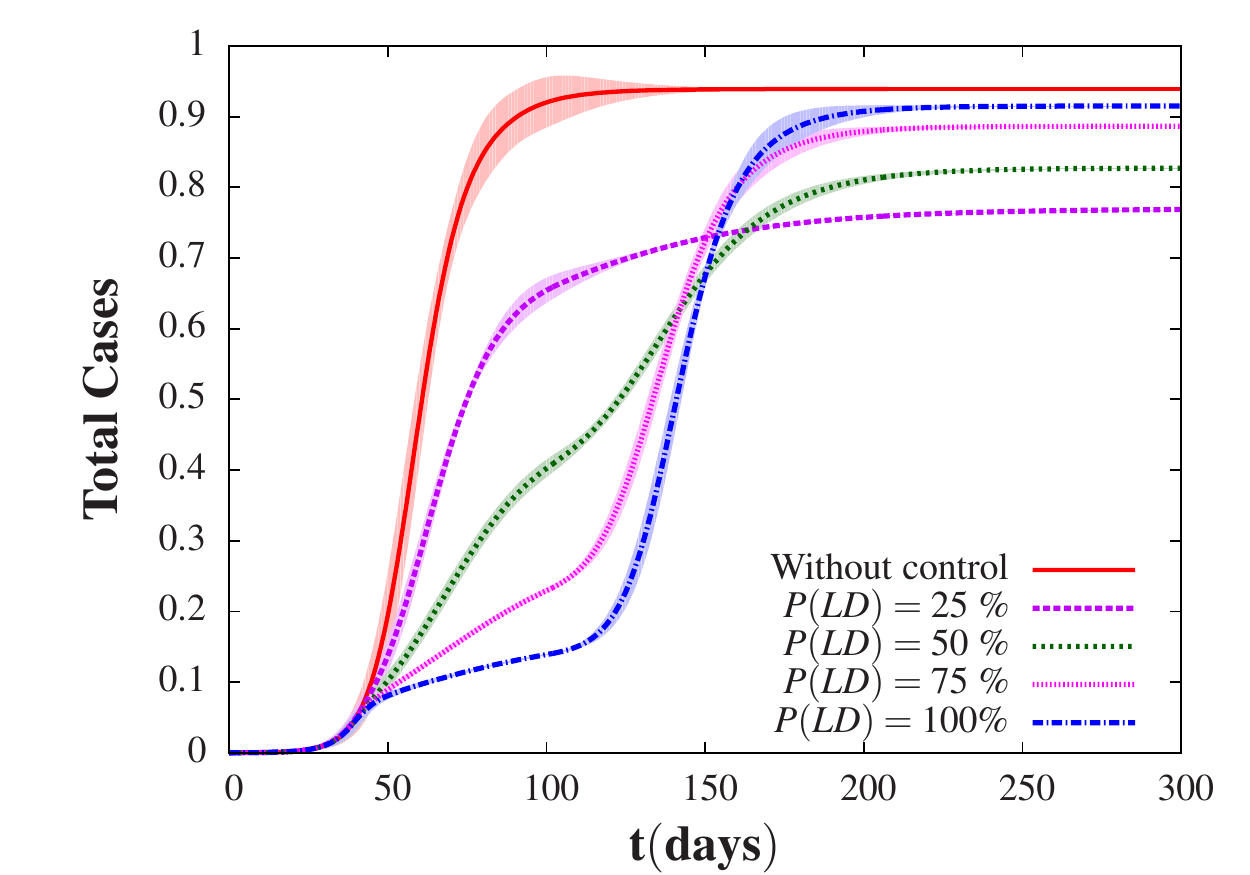}
% LDTCMp05.pdf: 0x0 px, 0dpi, nanxnan cm, bb=

            \caption{Lockdown started at $I(t) = 0.05$}
            \label{fig:lockdown_b}
    \end{subfigure}
        \begin{subfigure}[b]{0.5\textwidth}            
 \includegraphics[width=\textwidth]{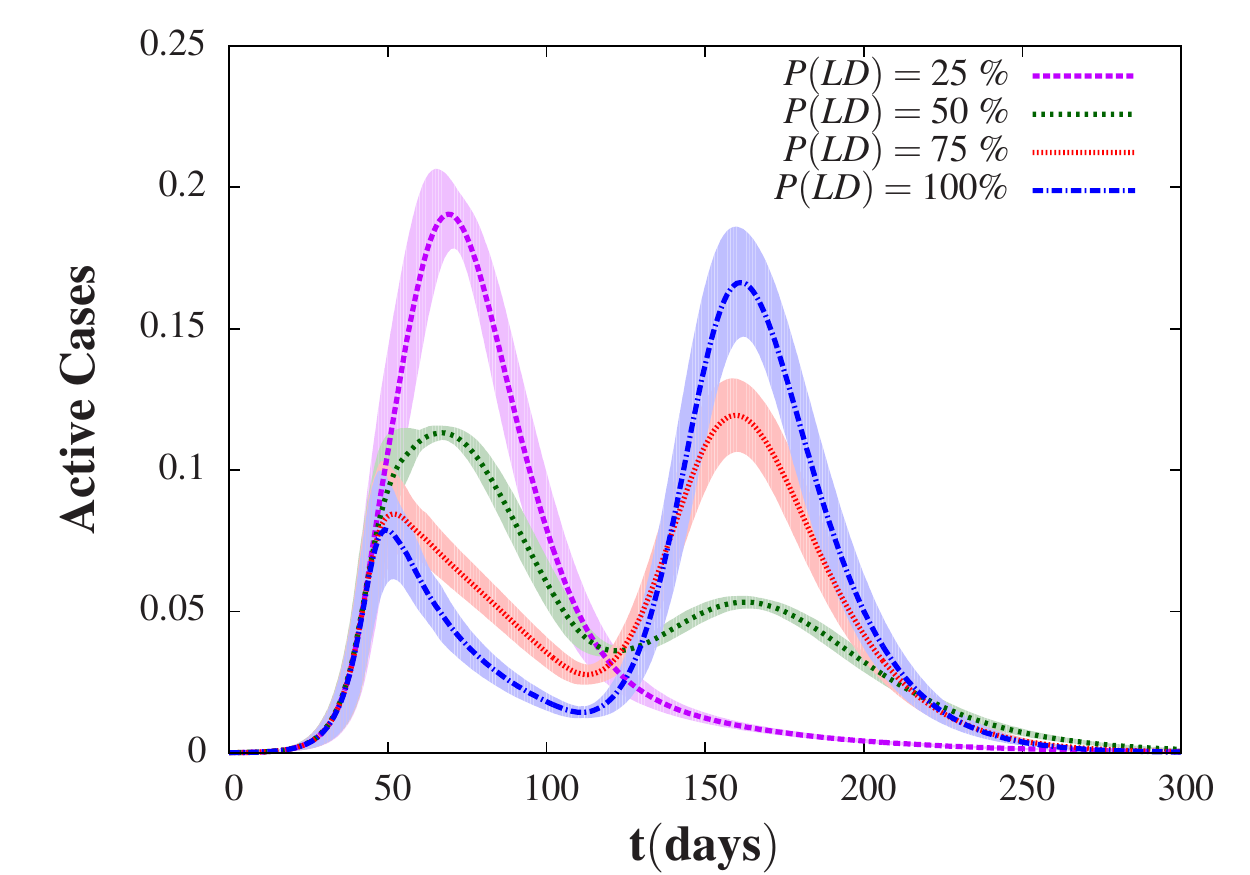}
 % LDInfp1.pdf: 0x0 px, 0dpi, 0.00x0.00 cm, bb=
            \caption{Lockdown started at $I(t) = 0.1$}
            \label{fig:lockdown_c}
    \end{subfigure}%
     %add desired spacing between images, e. g. ~, \quad, \qquad etc.
    \begin{subfigure}[b]{0.5\textwidth}
        \includegraphics[width=\textwidth]{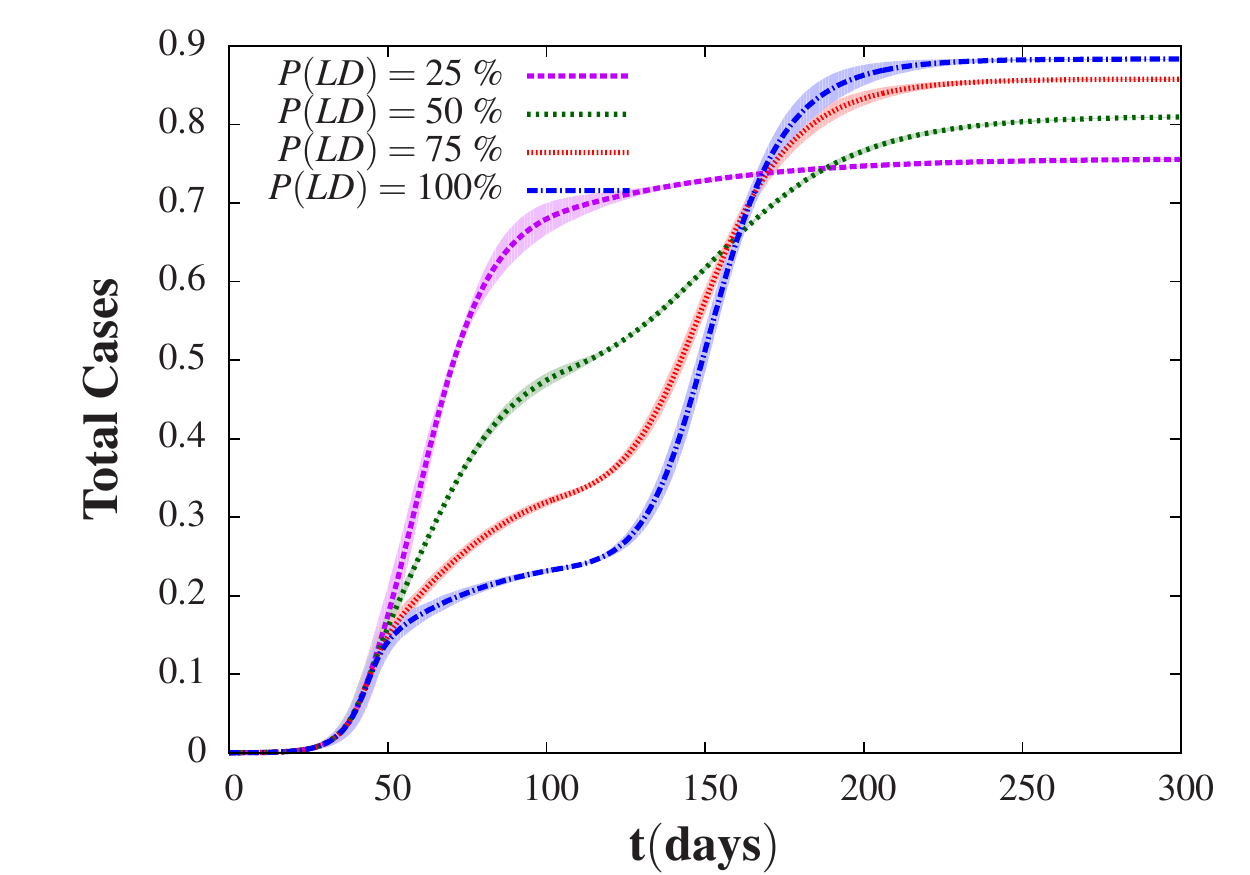}
            \caption{Lockdown started at $I(t) = 0.1$}
            \label{fig:lockdown_d}
    \end{subfigure}
    \caption{The evolution of active cases (left panels) and total cases (right panels) in the population when the spread is mitigated through partial and complete lockdowns. Different studies are performed by introducing lockdowns at different stages of an uncontrolled simulation -- $I(t) = 0.05$ (top panels) and $I(t) = 0.1$ (bottom panels). The lockdowns are maintained for the duration of 60-days, and the impact on the second outbreak is also explored after the lockdown is lifted. The shaded region show the standard deviation of the data form its mean value represented by lines.} 
    \label{fig:lockdown}
\end{figure}

The impact of lockdown on the spread of infection is examined in Fig.~\ref{fig:lockdown}. In the real world, the lockdown cannot be continued indefinitely, since social interactions are important for economic transactions as well as for the well-being of individuals. Therefore, a lockdown period of $60$ days is considered. There is always a risk of the second outbreak in the population after the lockdown is lifted, depending on the number density of the susceptible. The state of the system at this time is, therefore, important for understanding the further the spread and the final configuration attained by the system. The end state of the lockdown is strongly dependent on the time when this measure is introduced during the uncontrolled epidemic growth. In general, an early imposition will result a higher density of the susceptible who can get infected than a late imposition. Consequently, we examine two different scenarios by imposing the lockdown at two different times after an uncontrolled spread of the infection. In the first scenario (LD5P), the lockdown was imposed when the number density of the infected reached $5\%$ of the total population, whereas, in the second scenario (LD10P), it was imposed when the number density was $10\%$. An ideal implementation of the lockdown in which all individuals become isolated is difficult to achieve in the real population. For this reason, the cases studied here also consider the scenarios in which the fraction of the population in isolation varies. 

We first consider the LD5P cases. The lockdown is introduced at $t \approx 35$ days when the density of the infected reached about $5\%$ of the population, and maintained for another $60$ days. Here, we investigate the spread of the infection during and after the lockdown (Figs.~\ref{fig:lockdown_a} and \ref{fig:lockdown_b}). In the case with $100\%$ lockdown, in which everyone in the population isolates himself/herself, the density of the infectious stops growing immediately after the lockdown is imposed. Subsequently, it decreases with time as the infected nodes recover. After the $60^{th}$ day when the lockdown is withdrawn, the infectious start growing once again and attain a second peak because of the second outbreak. It is well known from earlier studies that a longer duration of the lockdown lowers the number of infectious \cite{Ferguson:2020, Bootsma:2007,Ruslan:2020,Prem:2020}, while the susceptible density remains almost the same; a sufficiently long lockdown can completely eradicate the infections in the system. Nevertheless, the system remains susceptible to a second outbreak if the susceptible are large in number. For a closed system completely eradicating the infectious is desirable; however, for systems in which the outside interactions are present, further mitigation efforts might be needed to keep the spread in check, as a small number of infectious coming in from the outside can lead to another outbreak. The severity of the second outbreak in such a system is investigated in detail below. 

The response of the system in partial lockdowns with probabilities $25\%$, $50\%$ and $75\%$ is also examined in Figs.~\ref{fig:lockdown_a} and \ref{fig:lockdown_b}. The overall behavior of the system in a partial lockdown varies from the uncontrolled response to that exhibited by the system in complete lockdown, depending on the probability of the implementation. For lockdown with $25\%$ probability, the number of active cases evolves similar to that with the uncontrolled simulation (Fig.~\ref{fig:lockdown_a}). There is no second outbreak, but the peak value of the active cases is slightly lower at approx $17\%$ of the total population compared to the uncontrolled case. Moreover, as shown in Fig.~\ref{fig:lockdown_b}, the total cases saturates at $78\%$ of the total population, as compared to $95\%$ for the uncontrolled simulation. The lockdown with $75\%$ probability, on the other hand, is closer to the complete lockdown case. After the lockdown is imposed, the active cases increases slightly before they start to fall, and there is a second peak after the lockdown is removed. The second peak forms at $20\%$ of the total population, a slightly smaller value compared to the complete lockdown case, in which the peak forms at $25\%$ of the total population. The reason is that more people get infected during the lockdown with $75\%$ probability, and the number of susceptible is therefore relatively lower when it is lifted. Considering the smaller second peak, the lockdown with $75\%$ probability is more desirable than the complete lockdown, if there is a second outbreak in the system. The saturation of the total cases with $75\%$ probability is close to that with the complete lockdown (Fig.~\ref{fig:lockdown_b}). The case with $50\%$ lockdown is the most interesting. In this case, two peaks of comparable strength develop for the active cases at about $10\%$ of the total population, and the total cases saturate at a value comparable to the case with $25\%$ probability. 

\begin{figure}[!ht]
\centering
    \begin{subfigure}[b]{0.5\textwidth}            
            \includegraphics[width=1\textwidth]{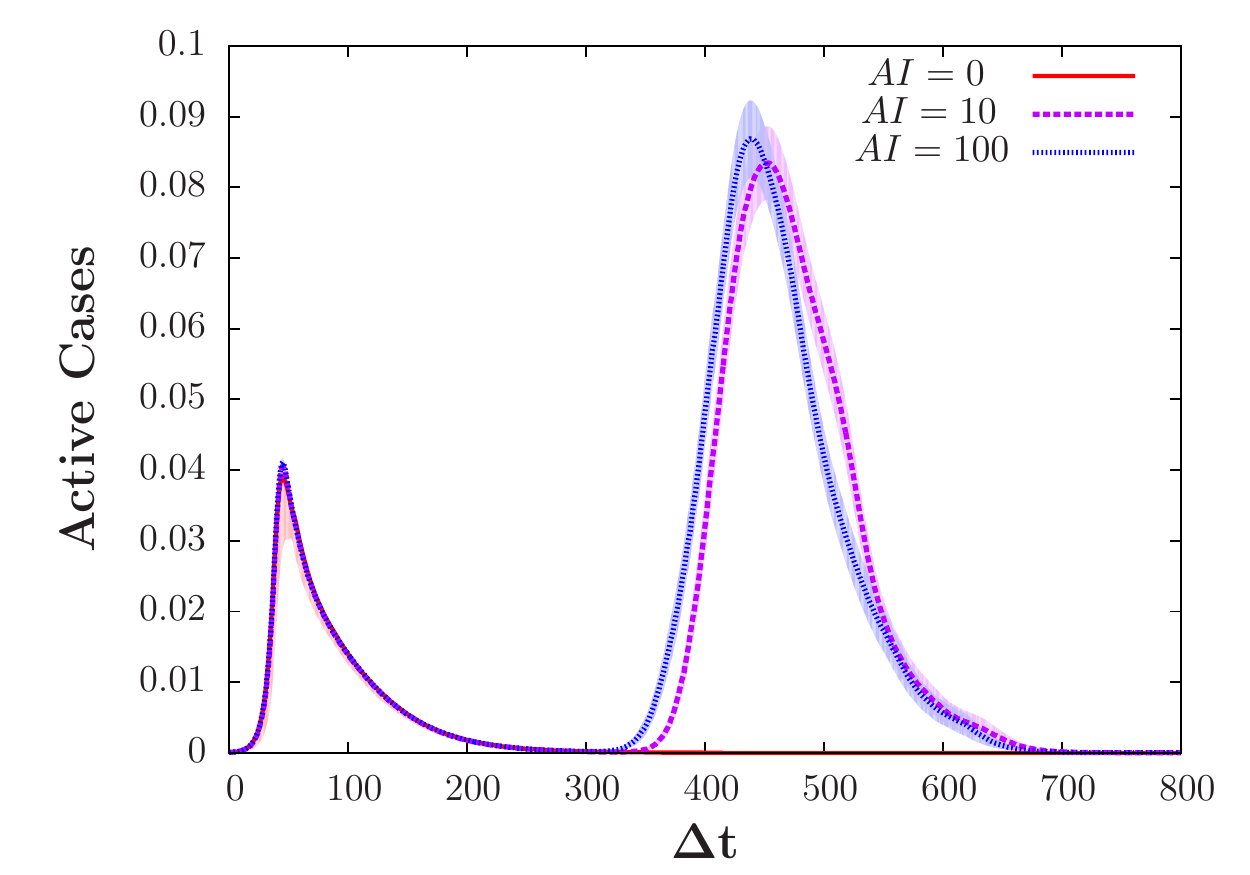}
            \caption{Complete lockdown started at $I(t) = 0.05$}
            \label{fig:lockdown_second_outbreak_a}
    \end{subfigure}%
     %add desired spacing between images, e. g. ~, \quad, \qquad etc.
    \begin{subfigure}[b]{0.5\textwidth}
            \centering
            \includegraphics[width=1\textwidth]{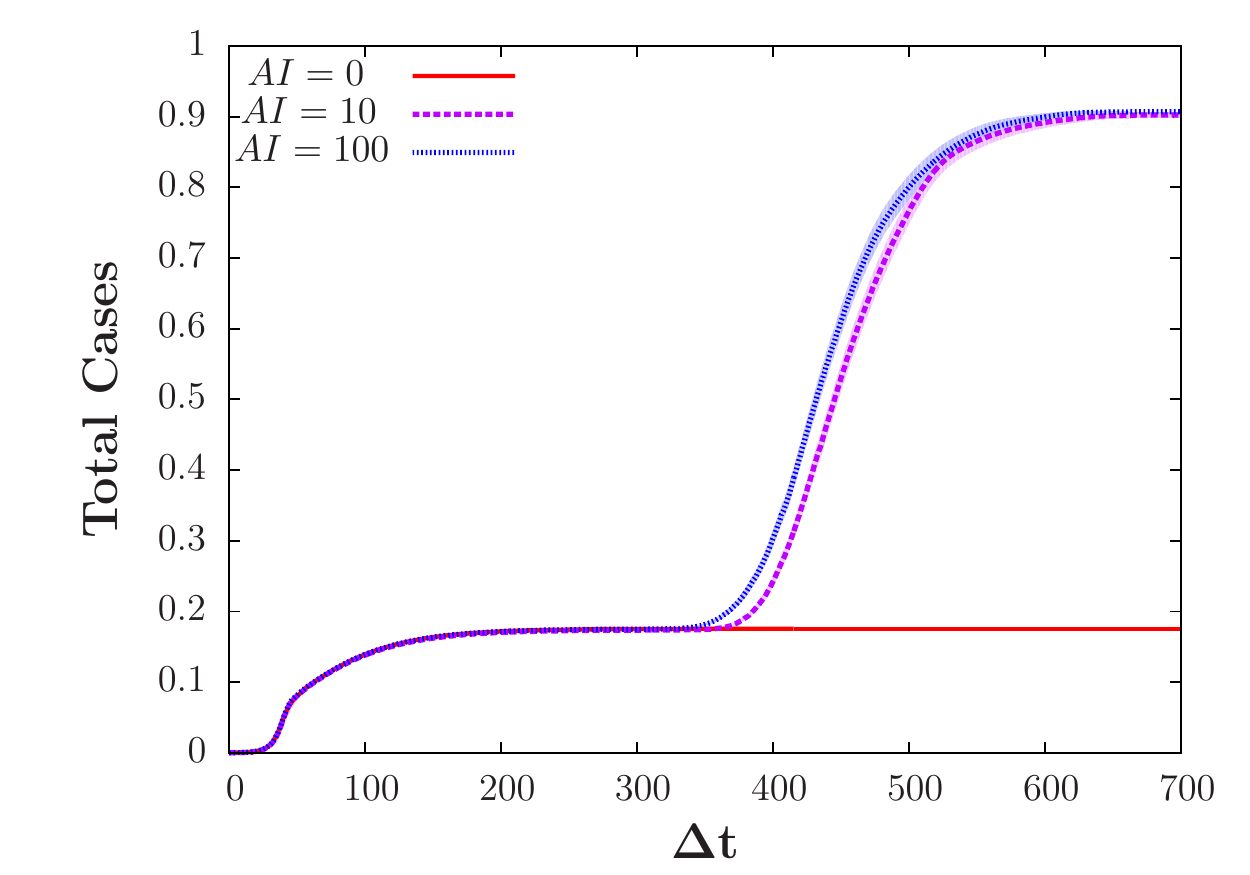}
            \caption{Complete lockdown started at $I(t) = 0.05$}
            \label{fig:lockdown_second_outbreak_b}
    \end{subfigure}
    
    \caption{Possibilities of second outbreak by introducing additional infectious at reopening after $300$ days of complete lockdown for $10^5$ number of nodes with $20$ mean connections. The reproduction number ${\cal R}_o=3.26$ and recovery period of $14$ days are used in the simulations and each curve is obtained after $100$ different realizations. The shaded regions show the standard error of the data form its mean value represented by lines.} 
    \label{fig:lockdown_second_outbreak}
\end{figure}

%This also concludes that there is a critcal threshold of the lockdown probability (below $50\%$) having a second peak which is also dependent on system configuration at the start of mitigation.     

It is crucial to notice that the second wave is proportional to the seriousness of the lockdown: the more effective the lockdown, the higher is the second peak since the number of susceptible increases at the end of the lockdown period. In the case with $25\%$ probability, the second peak is not observed because a large part of the population becomes infected by the time lockdown is lifted. The plot of the total cases generally shows three regimes: the first corresponding to the unmitigated growth, the second corresponding to lockdown, and the final corresponding to a converged configuration with herd immunity. When there is a second outbreak, there is an additional regime corresponding to the outbreak before the system transitions to the final configuration regime. 

 % In the 60-days lockdown case, $R(t)$ converges to a much lower value of about 0.24, whereas in the 40-days lockdown case $R(t)$ converges to about 0.85, which is much closer to the value observed in the uncontrolled simulation after reaching the herd immunity. As noted earlier, the difference mainly arises because of the small difference in the densities of the infectious at the start of the second breakout. 

%The results from $50\%$ lockdown cases are similar; however, some differences can be identified in the behavior of $I(t)$ and $R(t)$. For both 40- and 60-days lockdown cases, the first peak in $I(t)$ forms a few days after the lockdown is imposed. Also, the second peak is lower for the 40-days, but is higher for the 60-days case, in comparison to the corresponding cases with $100\%$ lockdown. Noticeable changes in $R(t)$ are also observed. Consistent with the observed $I(t)$, the value of $R(t)$ converges to about 0.85 for the 40-days case and to about 0.6 for the 60-days case, slightly higher values in comparison with the values attained in the corresponding $100\%$ lockdown simulations, but remain much smaller than the herd immunity value of the uncontrolled simulation.

Next, we consider the LD10P cases shown in Figs.\ref{fig:lockdown_c} and \ref{fig:lockdown_d}. The dynamics observed for these cases with complete and partial lockdowns imposed for 60-days are similar to those observed for LD5P cases. Some differences arise due to the larger number of the infectious at the start of the lockdown, resulting in a lower number of susceptible at the end of it. As a consequence, the peak in active cases curve during the second outbreak forms at slightly lower values, as compared to the LD5P cases. For example, in $50\%$ lockdown, the first peak occurs at $10\%$ of the population, but the second peak occurs at $5\%$ of the population, and the total cases saturates at $80\%$ of the population.  
%A departure from this trend is the case with the 60-days lockdown period. This case shows slightly higher peak value of $I(t)$ compared to the 60-days lockdown case from LD5P, although the number density of the susceptible is relatively smaller. The difference can be attributed to the larger number of the infectious at the beginning of the outbreak in the present 60-days case. The higher peak also correlates with the larger saturation value of about 0.7 for $R(t)$.    

From above results it is clear that lockdowns implemented with high probabilities can reduce the number of infectious during the time the measure is in place. Therefore, it is possible to get rid of the infection completely by imposing the lockdown over a sufficiently long time period. However, depending on the number density of the susceptible, the system remains at the risk of a second out break, after the lockdown is lifted and new infections are brought into the system from outside. It is likely that the system develops some resistance towards the second outbreak, since there are some recovered among the susceptible and starting the outbreak might require a large number of new infectious in the system. In Fig.~\ref{fig:lockdown_second_outbreak}, we investigate the relationship between the second outbreak and the number of new infectious in the system after complete lockdown is maintained for over 300 days. For this investigation, the LD5P case is considered, and the second outbreak is studied by introducing different numbers of infected ($0$ ,$10$ and $100$) at the end of the lockdown. Trivially, if no new infection is introduced into the system after the lockdown is lifted, the infection does not grow. This means that all the infectious in the system recover during the long lockdown. However, a second peak develops with $10$ and $100$ new infections are added in the system. As shown in Fig.~\ref{fig:lockdown_second_outbreak_a}, the active cases start increasing in both scenarios, but the growth is somewhat faster with $100$ new infections than $10$. The total cases saturates to about the same value ($90\%$ of the population), close to the value observed in the uncontrolled spread. Thus it is important to note that the effect of outside influences, although ignored in the closed systems, is important considered for the connected populations in the real world. At this point, we would like to emphasize that we can easily predict the second peak by equating the difference in the integral of the active cases after the lockdown with the total cases of the uncontrolled curve.

\subsection{Testing}
\label{Sec:Testing}
Testing is another effective strategy for handling the pandemic. This relies on our ability to identify the infectious in the population; once identified, they are isolated to stop the further spread. The mitigation through testing can be made more effective by tracing the source of the infection and the contacts of the infected. Without focusing on the details of how the infected are being identified, we assume to have a {\em testing mechanism} in place that identifies a fraction of the newly infected, defined as the effectiveness of testing, ${\cal{P(T)}}$. 
\begin{figure}[!ht]
\centering
    \begin{subfigure}[b]{0.5\textwidth}            
           \includegraphics[width=\textwidth]{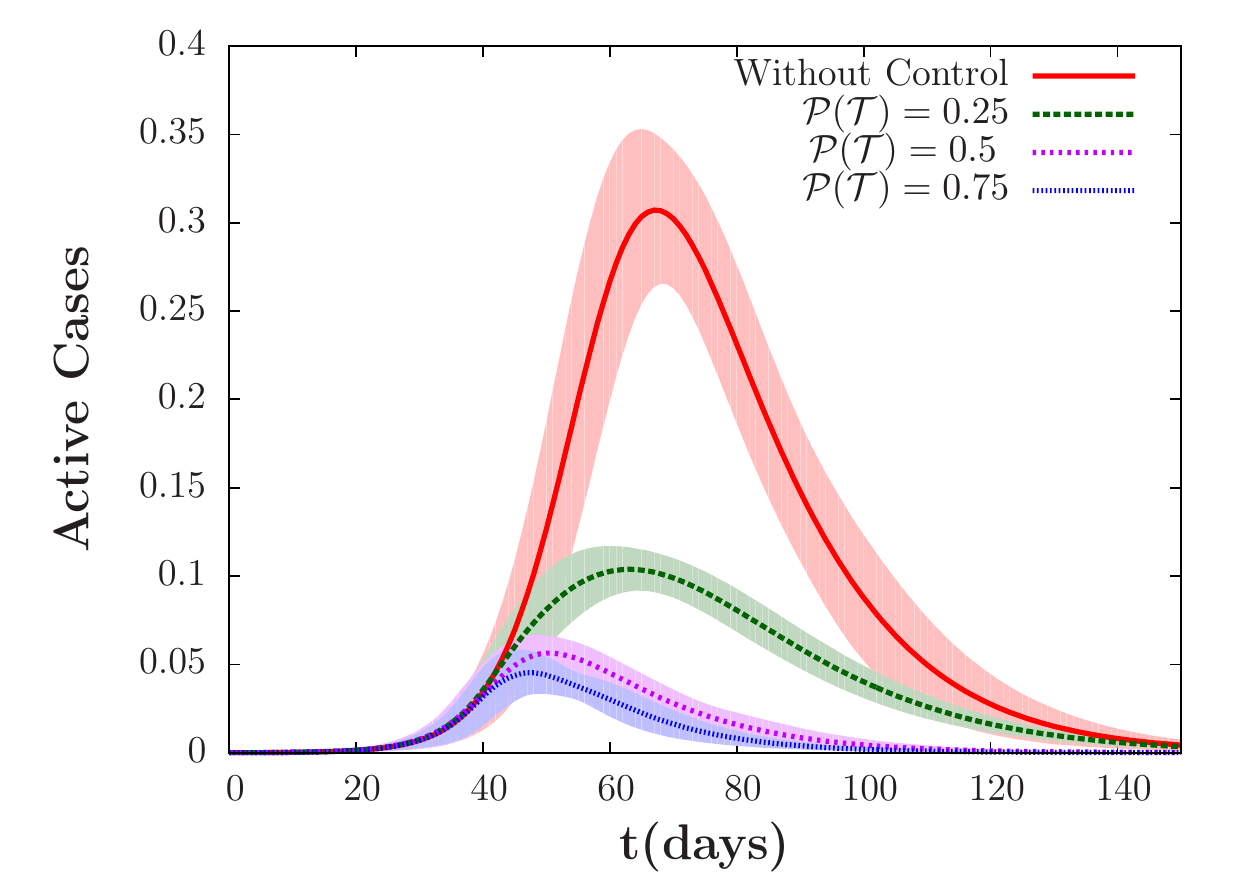}
            \caption{Testing started at $I(t) = 0.05$}
            \label{fig:testing_a}
    \end{subfigure}%    
    \begin{subfigure}[b]{0.5\textwidth}
            \centering
            \includegraphics[width=\textwidth]{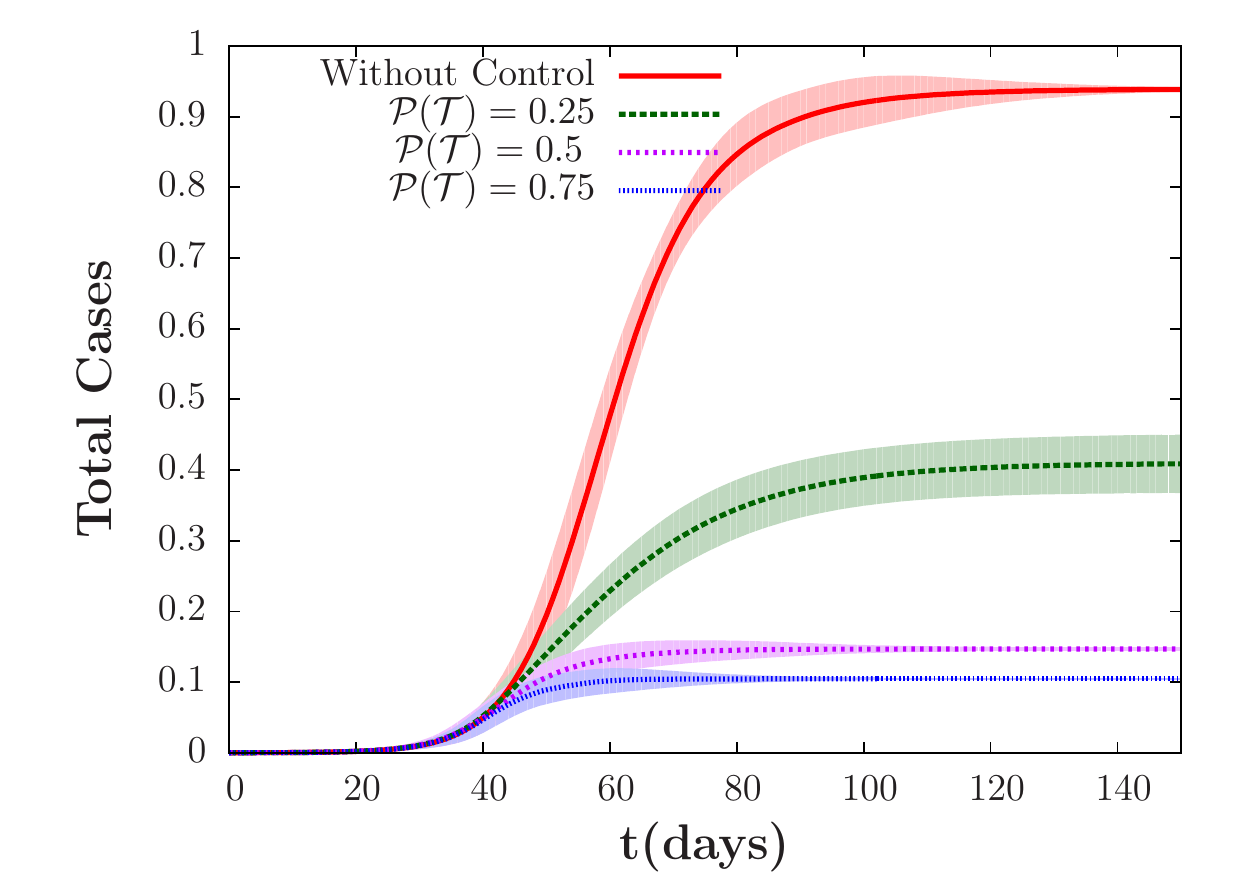}
            \caption{Testing started at $I(t) = 0.05$}
            \label{fig:testing_b}
    \end{subfigure}
        \begin{subfigure}[b]{0.5\textwidth}            
            \includegraphics[width=\textwidth]{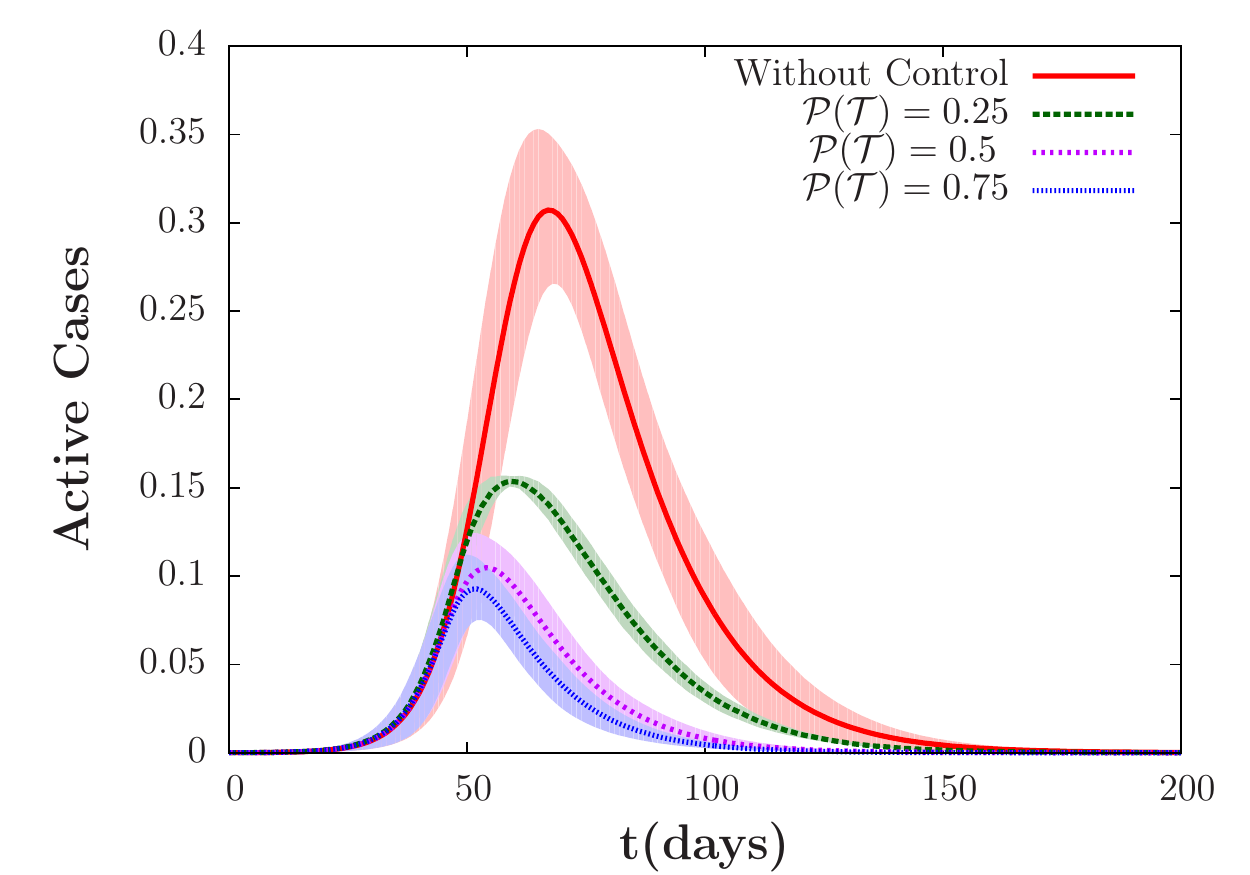}
            \caption{Testing started at $I(t) = 0.1$}
            \label{fig:testing_c}
    \end{subfigure}%
     %add desired spacing between images, e. g. ~, \quad, \qquad etc.
    \begin{subfigure}[b]{0.5\textwidth}
            \centering
            \includegraphics[width=\textwidth]{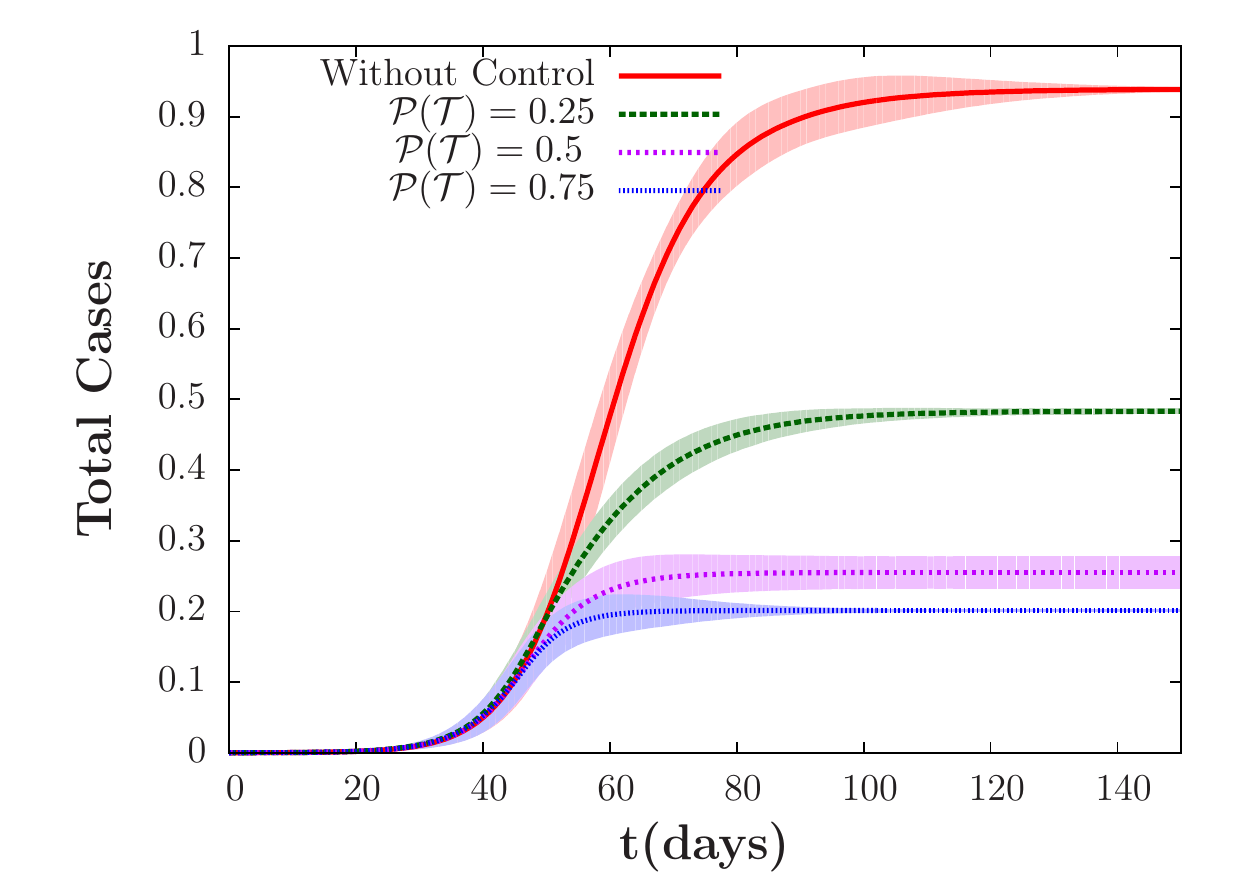}
            \caption{Testing started at $I(t) = 0.1$}
            \label{fig:testing_d}
    \end{subfigure}
    \caption{The evolution of the active cases $I(t)$ (left panels) and the total cases $I(t) + R(t)$ (right panels) in the population when the spread of infection is controlled through testing of varying effectiveness. Two different studies are performed employing testing measures when density of $I(t) = 0.05$ and $0.1$. For comparison, the active and the total cases from the uncontrolled simulation are also plotted with solid lines. The shaded regions show the standard deviation of the data form its mean value represented by lines}
    \label{fig:testing}
\end{figure}
In the random network, the effect of testing is captured by randomly selecting an infected node with a probability determined by the effectiveness of the testing, and then isolating the node by blocking any further transmission through contacts. We have also included a time delay, $T_{in}$, after which the infected nodes become available for testing. The time delay takes into account the incubation time of the infection. However, during this time an infected node continues to spread the infection. In this study, $T_{in} = 2 \, \rm{day}$ is used, corresponding to the incubation time of COVID-19. The time evolution of the active and the total cases in the population with testing measures of varying effectiveness are shown in Fig.~\ref{fig:testing}. Similar to the study of lockdown, two scenarios are considered that differ form each other in terms of the start of testing. Cases TS5P correspond to those simulations in which testing is started after the infectious reached $5\%$ of the total population, whereas cases TS10P correspond to those in which testing is started at $10\%$ of the infected density in the population. 

\begin{figure}[!ht]
\centering
    \begin{subfigure}[b]{0.5\textwidth}            
          \includegraphics[width=\textwidth]{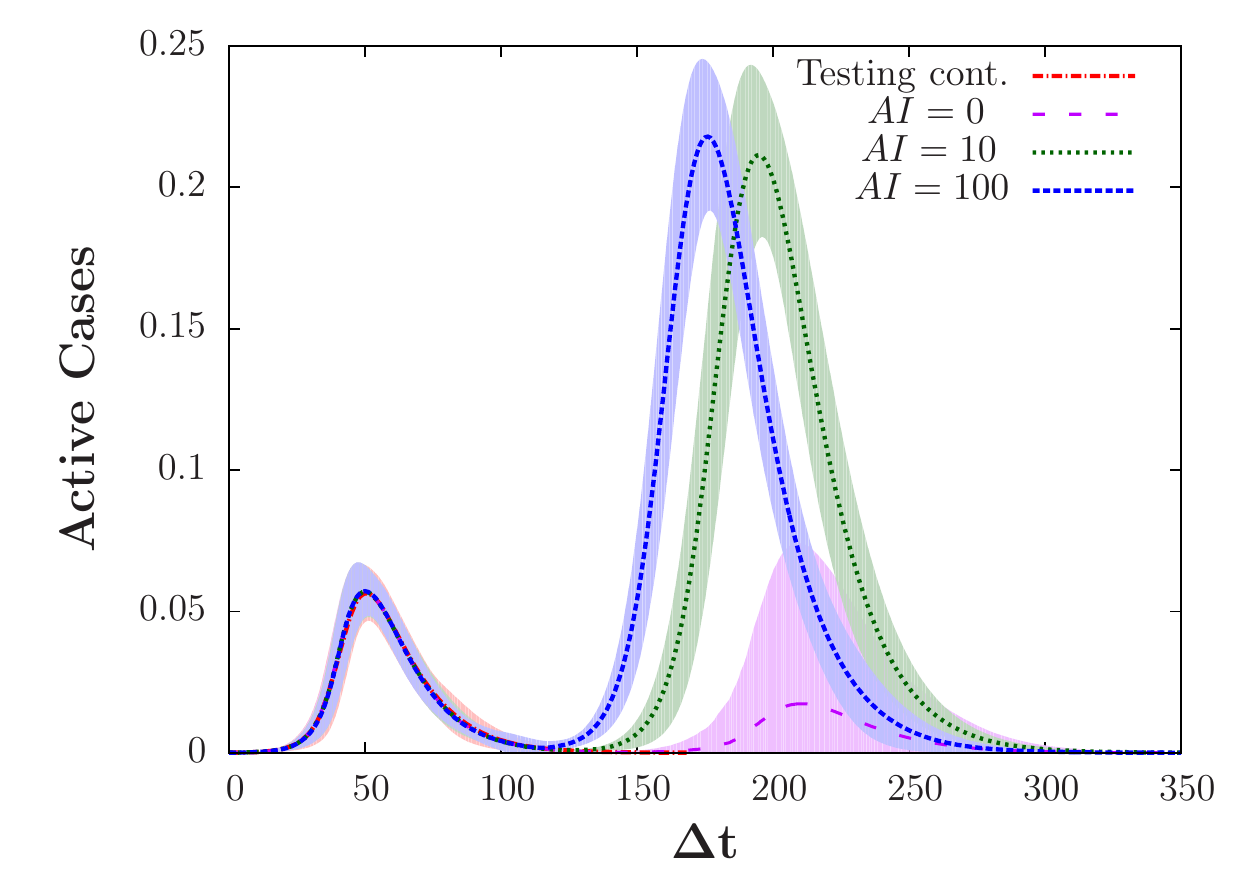}
% AIInftestP5.pdf: 0x0 px, 0dpi, nanxnan cm, bb=
            \caption{Testing stopped at $I(t) = 0.001$}
            \label{fig:AItesting_a}
    \end{subfigure}%
    \begin{subfigure}[b]{0.5\textwidth}
            \includegraphics[width=\textwidth]{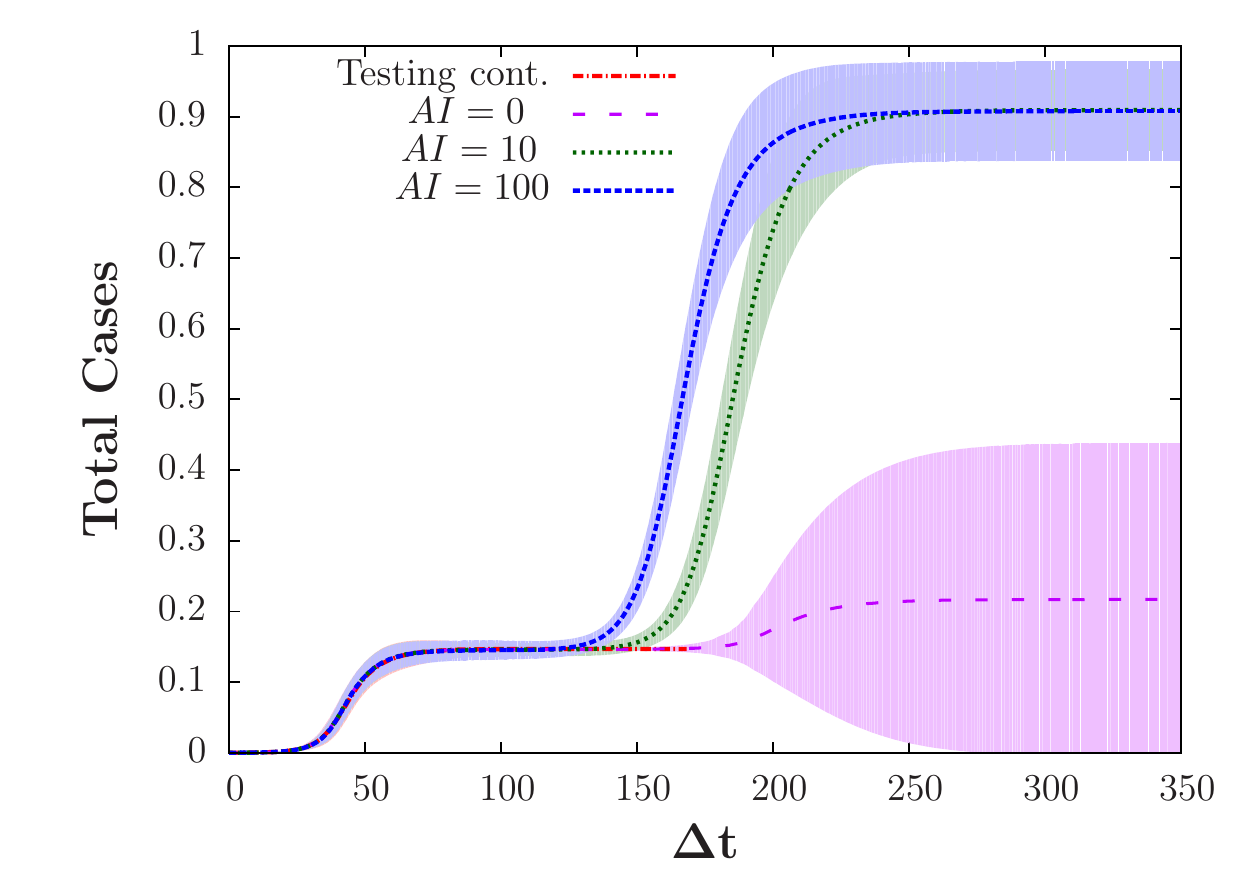}
% AIRectestP5.pdf: 0x0 px, 0dpi, nanxnan cm, bb=

            \caption{Testing stopped at $I(t) = 0.001$}
            \label{fig:AItesting_b}
    \end{subfigure}
        \begin{subfigure}[b]{0.5\textwidth}            
          \includegraphics[width=\textwidth]{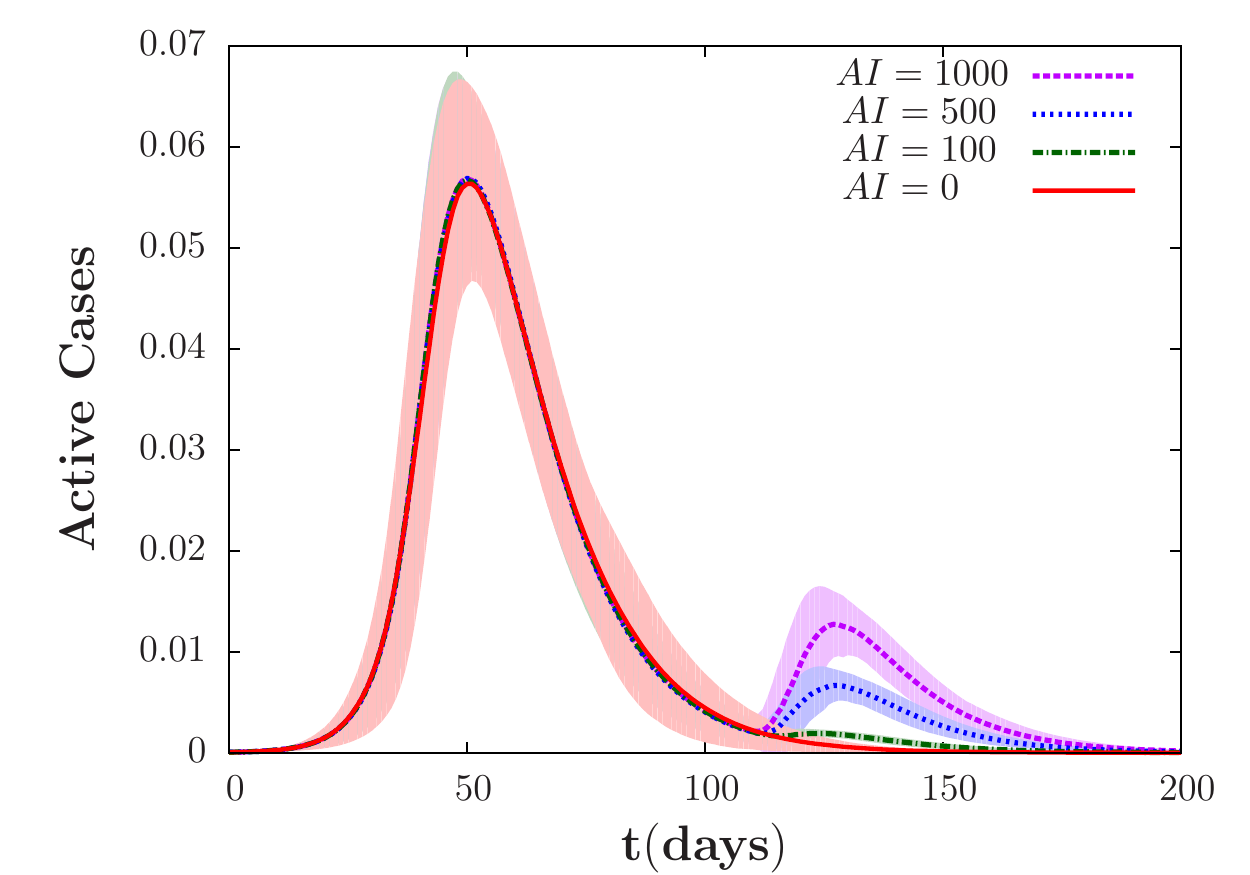}
% SAIInftestP5.pdf: 0x0 px, 0dpi, nanxnan cm, bb=

            \caption{Testing continue}
            \label{fig:AItesting_c}
    \end{subfigure}%
     %add desired spacing between images, e. g. ~, \quad, \qquad etc.
    \begin{subfigure}[b]{0.5\textwidth}
            \includegraphics[width=\textwidth]{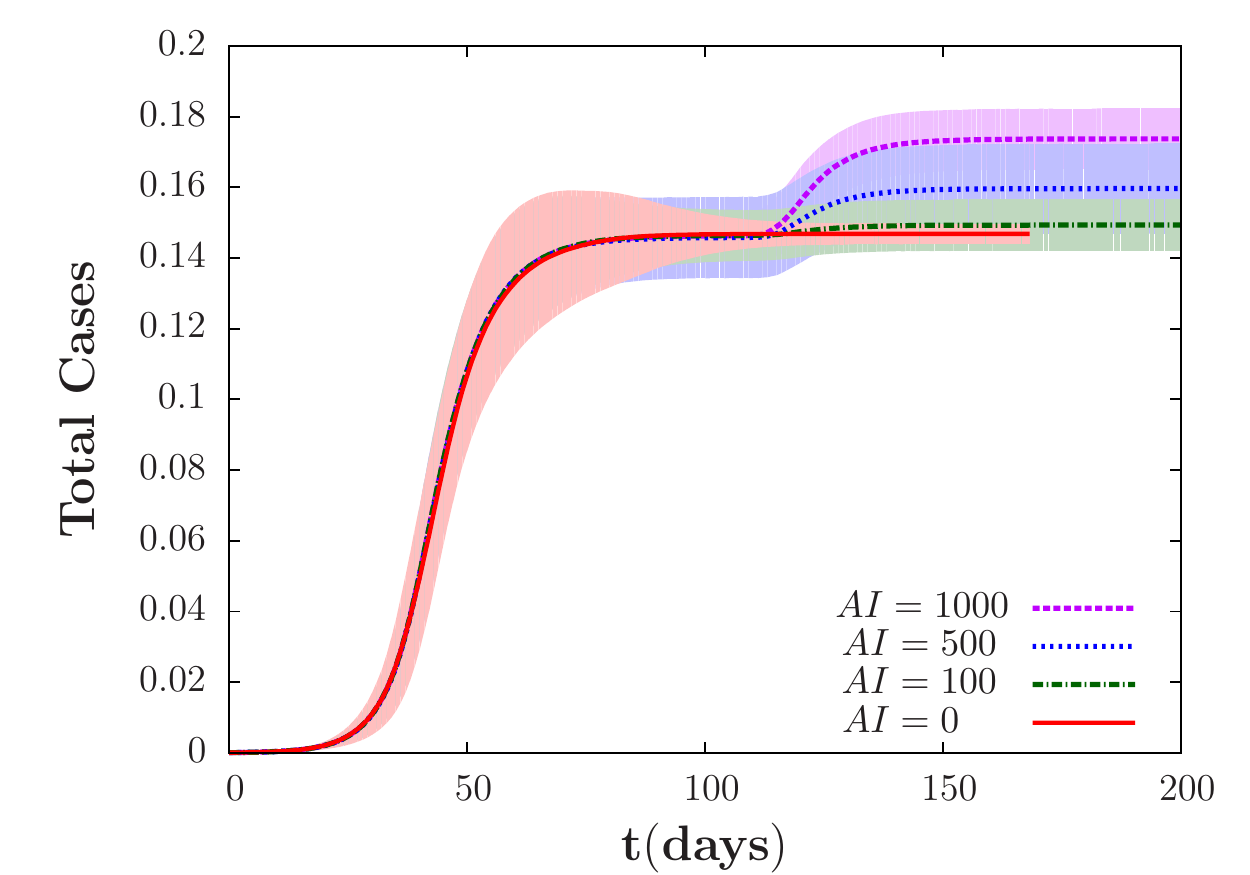}
            \caption{Testing continue}
            \label{fig:AItesting_d}
    \end{subfigure}
    \caption{The evolution of the active cases $I(t)$ (left panels) and the total cases (right panels) in the population when the spread of infection is controlled through testing with effectiveness,{\cal{P}(T)} of $50\%$. Two different studies are performed: the upper panels when testing is stopped at $110^{th}$ day and $10$ and $100$ new infectious are introduced to capture the external influences, and the lower panel when testing is continued at all times and $10$, $100$ and $1000$ new infectious are introduced at $110^{th}$ day. The parameters used for the simulation are $N=10^5,~\alpha=0.25,~\alpha=1/14$ and averaged over $500$ different realizations. The shaded regions show the standard deviation of the data form its mean value represented by lines.}
    \label{fig:AItesting}
\end{figure}

First, we consider the TS5P cases (Fig.~\ref{fig:testing_a} and \ref{fig:testing_b}). Figure ~\ref{fig:testing_a} depicts the effect of testing of varying effectiveness (${\cal{P(T)}}$) on the active cases. It can be observed from the figure that testing is capable of suppressing the peak value, and it becomes more efficient with higher effectiveness. Notice that with effectiveness of $75\%$, the peak in the active cases is only slightly higher than $0.05$. Even with effectiveness of $50\%$, the testing can reduce the number of active cases at the peak by $2/3$ compared to the value attained in the uncontrolled system. An examination of the total cases in Fig.~\ref{fig:testing_b} shows smaller saturation values compared to the uncontrolled simulation, depending on the effectiveness of testing. For example, with effectiveness of $75\%$, the total cases asymptote to a value close to $0.2$, i.e., only $20\%$ of the population get infected at the end. Even in the worst testing measure with $25\%$ effectiveness, the total cases saturate to a much lower value of $50\%$ than the uncontrolled simulation with $90\%$ infected. It is worth noting that with the value of $R_0$ chosen for the study, the heard immunity in an uncontrolled growth is achieved after $90-95\%$ of the population become infected. Thus, there is a possibility of a second wave if the testing is stopped and if there is an influx of new infection into the system. This aspect of the problem is further investigated in Fig.~\ref{fig:AItesting}.

The results obtained from TS10P simulations are shown in Figs.~\ref{fig:testing_c} and \ref{fig:testing_d}. The resemblance between the results from TS10P simulations and those from TS5P simulations are striking. However, small differences can be identified, arising mainly due to different configurations at the time testing was introduced. In TS10P cases, testing started when the active cases and the recovered were larger compared to that in TS5P cases. 
The longtime convergence of the total cases to lower values in different TS5P and TS10P simulations compared to the uncontrolled simulation suggests smaller effective ${\cal R}_e$ for the system with testing. The dynamics are likely to revert back to that of the uncontrolled system without testing, adjusting $R_e$ to higher values. This suggests a second outbreak in the system, especially when there is a high influx of infection from the outside. This scenario is further investigated in Fig.~\ref{fig:AItesting}, where the evolution of the active and the total cases are plotted in the late-time configuration of the TSP5 case with $50\%$ effective testing. In this simulation, we infect some susceptible in different realizations when the active cases becomes $100$ (close to $100$th day of the spread) after the peak. As with the system under lockdown, the second outbreak is studied by changing the number of infectious, marking the beginning of the second outbreak. The peak in the active cases curve increases as the number of new infectious increases from $0$ to $10$ when testing is stopped, as shown in Fig. \ref{fig:AItesting_a}, and further saturates at the same level with any increase in the new infectious. Examining the total cases curves in Fig~\ref{fig:AItesting_b} reveals that with $10$ and $100$ new infections, total recovered asymptotes to a value close to $90\%$ of the total population, similar to what was achieve in the uncontrolled system. Thus, it is important to note that the system is vulnerable to new infections and, therefore, the mitigation should be continued to stopped the second out break. The effect on the continued testing scenario on the added infections (100, 500 and 1000) has also been explored in Figs.~\ref{fig:AItesting_c} and \ref{fig:AItesting_d}. It can be seen from the figures that with continues testing, the added infection does not result into a prominant second outbreak, and the infection quickly gets suppressed, while the total cases increase marginally. Even adding $1000$ infections, the testing with ${\cal P(T)}=50\%$ works well to keep the second peak in control.

\section{Analysis of COVID-19 Statistics of South Korea, Germany and New York}
\label{Sec:Realdata}
In earlier sections, we have discussed how the lockdown and testing strategies can be implemented using a random network, simulating the SIR dynamics. Utilizing the developed models, the dynamics of pandemic growth is studied by varying the probability of implementing lockdown and changing the effectiveness of testing. However, the predictive capabilities of the model in capturing the growth of an epidemic in the real population is not clear, as the model tries to incorporate only the most essential features of the real system while ignoring many of inherent complexities such as heterogeneity in the population density, non-uniformity in contacts, and other spatio-temporal variabilities. Here, we demonstrated that the model, although simple, can be tuned to imitate the behavior of a real system. We demonstrated the applicability by simulating the spread of COVID-19 in three different regions: South Korea, Germany and New York and show that the epidemic growth predicted by our model agrees with the real data. With this study we also realize how to fine-tune and robust the model parameters are to achieve a trend of the infection growth in realistic model. This comparison also help us understand the implications and needs of the lockdown and testing in realistic systems to bring the spread of the epidemic under control and maintaining the infectious to a low level. 

The random network used in these studies employs one million nodes. Although the number of individuals in the countries studied here can be up to 90 times larger, with one million node the network provides a good representation of the population as the number of cases are relatively small, only few hundred thousand. The average number of connection per node is $40$, which is reasonable considering the restrictions on travel and mass gatherings in these countries. We have also used $20$ mean connections per node, and similar results were obtained after modifying $\alpha$ by a small magnitude. The average connectivity between the nodes has an effect on the transmission rate is well know for discrete systems and this characteristics sets them apart from the continuous SIR model~\cite{Keeling:2005}. In all the three countries, we choose the effective infection rate between $0.4-0.35$. The $\alpha$ values are obtained by fitting the SIR predictions of the active cases with the real data during the early phases of the spread. Considering the recovery rate of $\gamma = 1/14 \, \rm{day}^{-1}$, corresponding to the mean recovery time of $14$ days for COVID-19, the reproduction number of these simulations are ${\cal R}_o = 4.9-5.6$. For these ${\cal R}_o$ values, the mean-field system achieves herd immunity after the infection spread to almost the entire population. In a real system, however, the herd immunity will be achieved at a smaller fraction since the mean-field always overestimates the number of total cases. We implement both lockdown and testing in these studies. The lockdown is implemented corresponding to the date when stay at home order was released by the states/countries. The testing period is approximated by observing the trend of the active cases and using information announced in the news broadcast. Further, we have tuned the parameters of the lockdown probability and the testing effectiveness to best capture the trend in the epidemic growth.
\begin{figure}[!ht]
\centering
    \begin{subfigure}[b]{0.5\textwidth}            
     \centering
     \includegraphics[width=\textwidth]{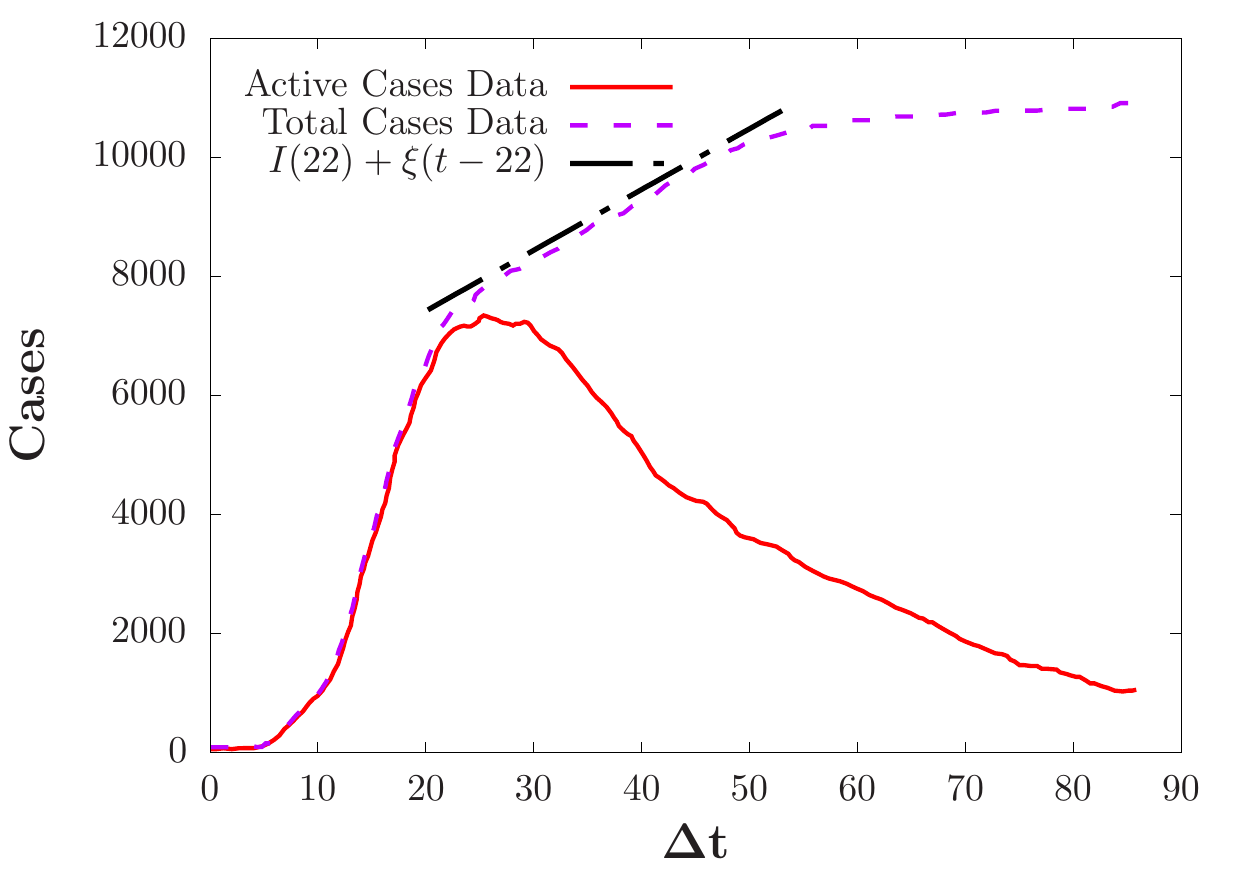}
   \caption{South Korea COVID-19 data}
    \label{SK_data} 
    \end{subfigure}%
     %add desired spacing between images, e. g. ~, \quad, \qquad etc.
       \begin{subfigure}[b]{0.5\textwidth}
       \centering
       \includegraphics[width=\textwidth]{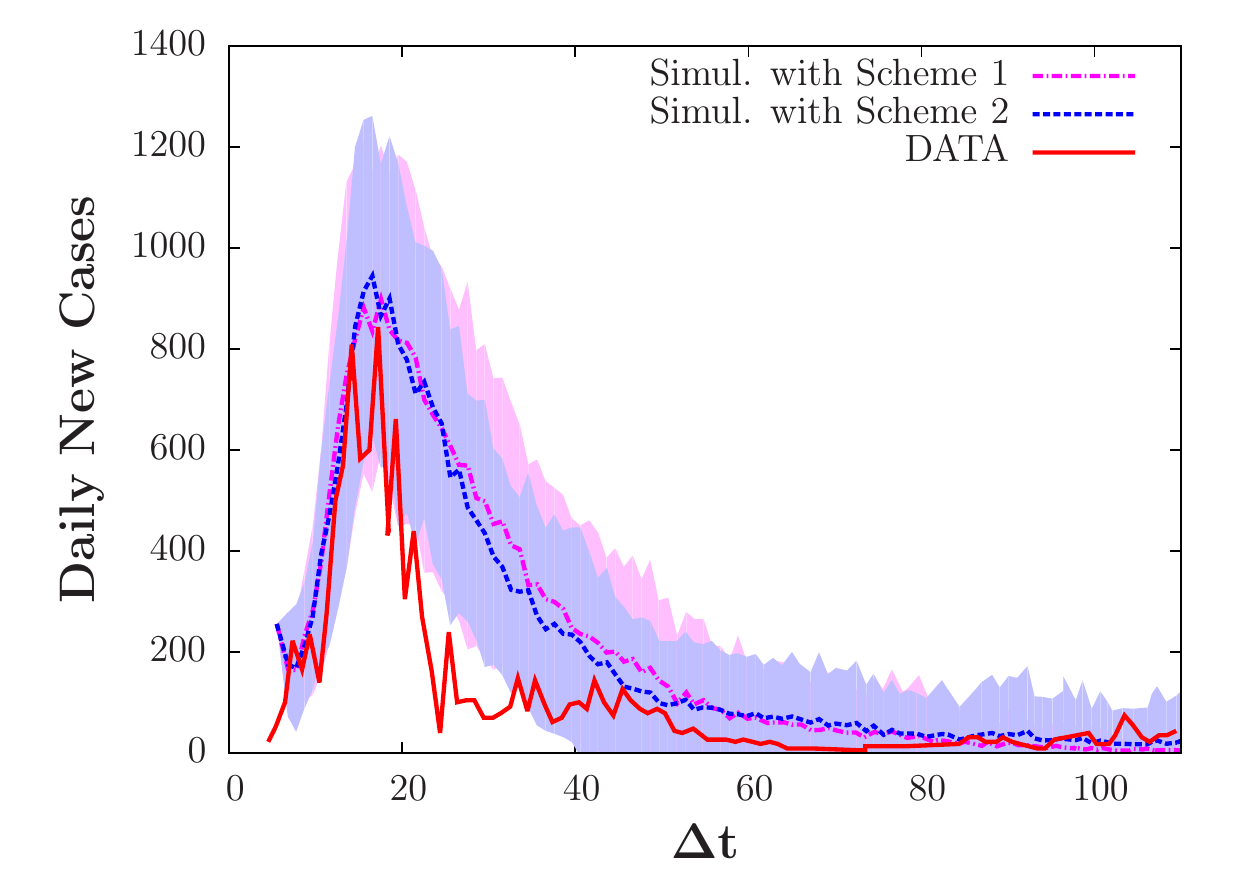}
% TCSK.pdf: 0x0 px, 0dpi, 0.00x0.00 cm, bb=
    \caption{South Korea daily new cases}
     \label{DCSK}
     \end{subfigure}
     \begin{subfigure}[b]{0.5\textwidth}            
     \centering
     \includegraphics[width=\textwidth]{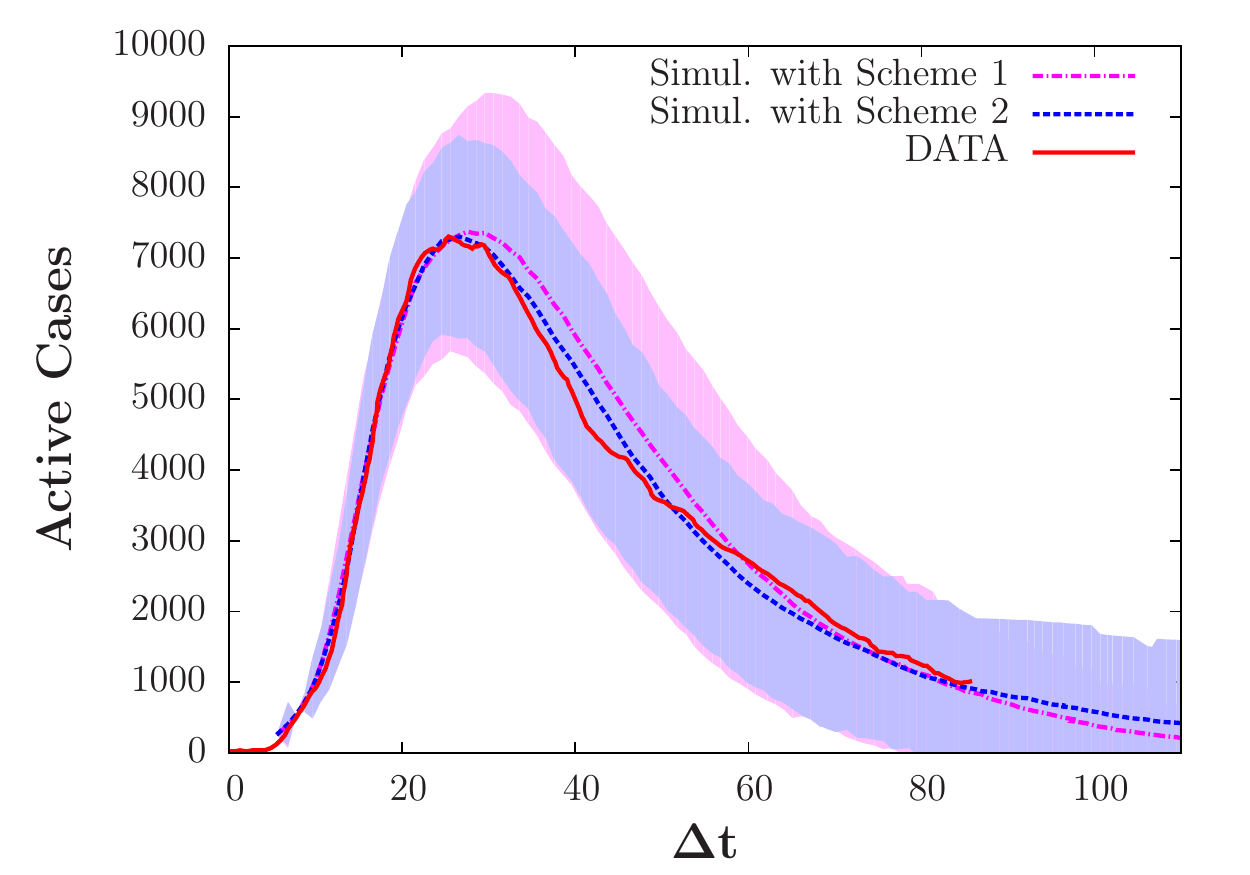}
   \caption{South Korea active cases}
    \label{ACSK}
    \end{subfigure}%
     %add desired spacing between images, e. g. ~, \quad, \qquad etc.
       \begin{subfigure}[b]{0.5\textwidth}
       \centering
       \includegraphics[width=\textwidth]{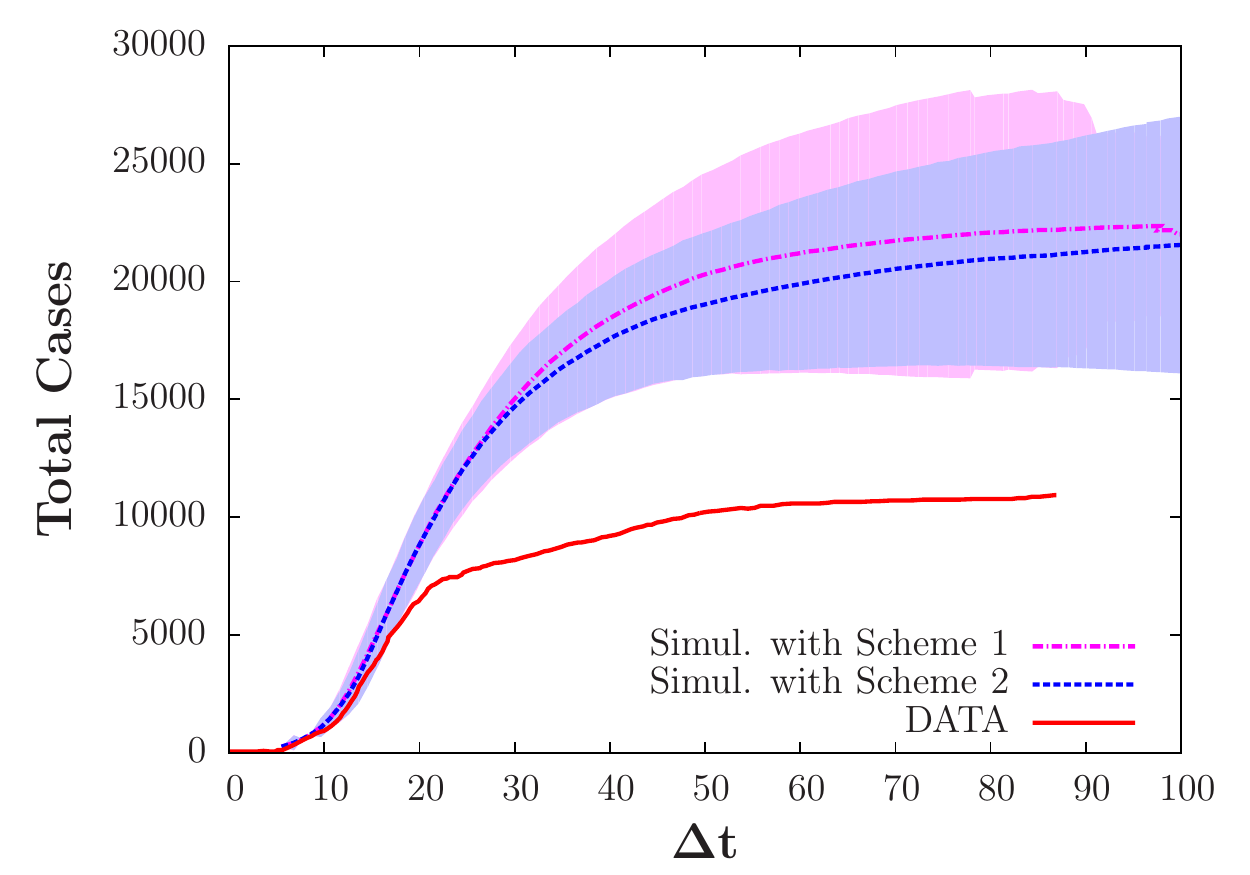}
% TCSK.pdf: 0x0 px, 0dpi, 0.00x0.00 cm, bb=
    \caption{South Korea: Total cases}
     \label{TCSK}
     \end{subfigure}
    \caption{Comparison plot of South Korea COVID-19 epidemic data with our simulation results obtained from SIR model over a random graph. In Fig. \ref{SK_data}, the linear trend in the total cases is the time after drive through testing started by South Korea. The data compared here is from {\em $15th$ February 2020}. The random graph used have one million nodes with $40$ mean edges and averaged over 100 different realizations. The other parameters $\alpha=0.4$, $\gamma=1/14$ are used. The shaded regions represent the standard deviation with its mean value denoted by lines.} 
    \label{SK_result}
\end{figure}

\subsection{South Korea}
\label{SouthKorea}
Earlier studies have estimated the intrinsic growth rate of COVID-19 in South Korea (SK) as $\alpha= 0.6 \, \rm{day^{-1}}$ \cite{SHIM:2020}. However, we have calculated $\alpha$ by fitting the data of the active cases with the SIR prediction for the first ten days and using recovery rate $\gamma=1/14 \, \rm{day^{-1}}$. The value obtained by best fit for simulation started with ten initial infection ($I(0)=10$) is $\alpha \approx 0.4$. The recovery rate can also be approximated from data by calculating the time when the total cases and the active cases start deviating, which is around $16th$ days from the day data is provided in Worldometer. We start the simulation at the time when the number of infectious in the population is $10$, i.e., $I(0)=10^{-5}$.

We simulated our modified model that contains testing as well as lockdown (Social distancing) strategies, as explained in Sec.~\ref{Sec:Testing} and \ref{Sec:Lockdown} and compare the results with SK statistics. We start the simulation without any mitigation, and implement partial lockdown with a small probability of $15\%$ on $23rd$ February, when the number of active cases is $I(t)=600$. Later, testing with effectiveness value of $65\%$ is started on $2^{th}$ of March, when the drive through testing began in SK with the number of infected reaching $4300$. The lockdown probability and the testing effectiveness are chosen to capture the trend in active cases accurately. As shown in Fig.~\ref{ACSK}, the simulated curve for the active cases compares well with the real data and is within the standard deviation. However, there is a small deviation between the two curves during the period when the total cases show a linear regime, resulting from a rather sharp change in the slope of the active cases from mid-March to early April. This trend has been difficult to capture in our simulation. This sharp change is of 30-days long and can be an artifact of a more effective implementation of the mitigation strategies, possibly with time-dependent variations. During this period, the number of infections per day is low, at around $100$ infection per day, which can be seen in the daily cases in Fig.~\ref{DCSK} \cite{Worldometer}. We also note that the present study does not take into account the asymptomatic cases who spread the infection, but are underrepresented in the data, and this can also cause serious problems in capturing the trend of the real system.  

We try to improve the prediction of the model described ({\em scheme 1}) during the linear regime of the total cases by changing the testing effectiveness to $70\%$ ({\em scheme 2}); thus illustrating the model sensitivity to this parameter. With this value of testing effectiveness, the linear regime is well represented, but the late time behavior under-represents the active cases (not shown). In order to rectify this problem, the effectiveness parameter is changed again to $60\%$ after 30-days. In scheme 2, we are able to obtain a linear regime in the total cases, but its start is delayed to $40th$-day, as compared to $20th$-day in the real data (Fig.~\ref{TCSK}). Moreover, the total cases curve rises as much as twice the magnitude of the real data. 
 
Finally, we note that a large number of individuals in the population remains susceptible to a second outbreak. Therefore, continuing testing with same effectiveness will be desirable for avoiding the second wave.

\subsection{Germany}
\begin{figure}[!ht]
    \begin{subfigure}[b]{0.5\textwidth}            
    \includegraphics[width=\textwidth]{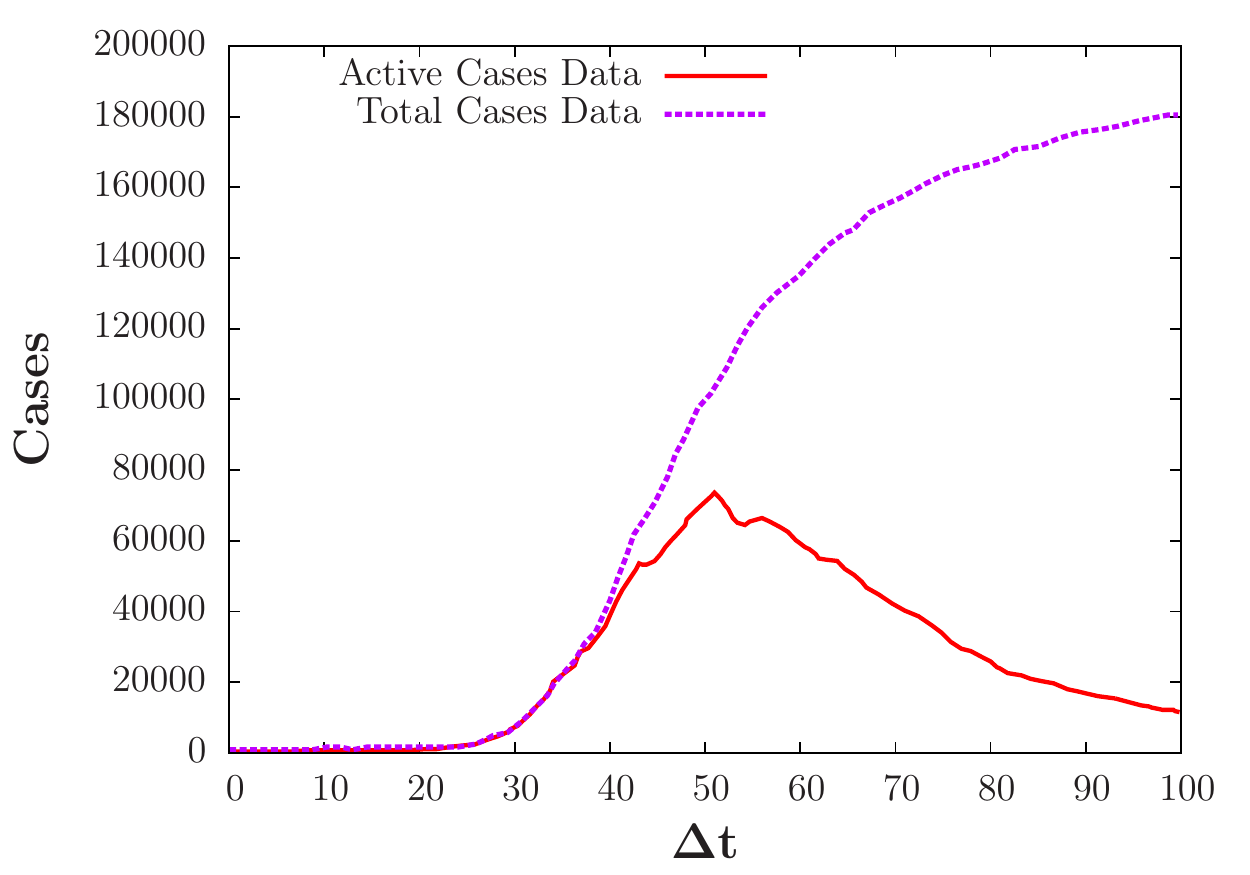}
% GD.pdf: 0x0 px, 0dpi, 0.00x0.00 cm, bb=
    \caption{Germany COVID-19 data}
   \label{G_data}
\end{subfigure}
\begin{subfigure}[b]{0.5\textwidth}            
    \includegraphics[width=\textwidth]{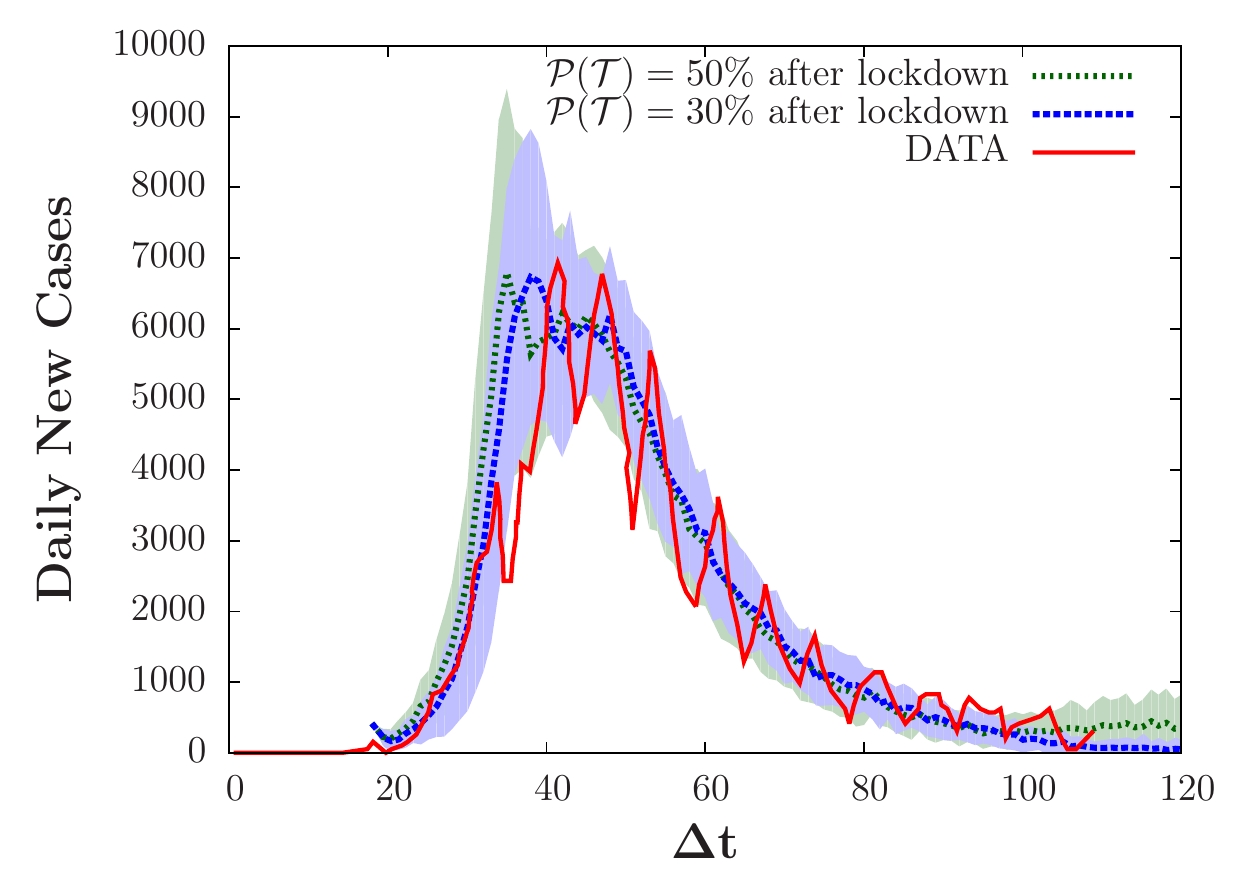}
     \caption{Germany daily infected case}
    \label{fig:DCGER}
\end{subfigure}
\begin{subfigure}[b]{0.5\textwidth}            
    \includegraphics[width=\textwidth]{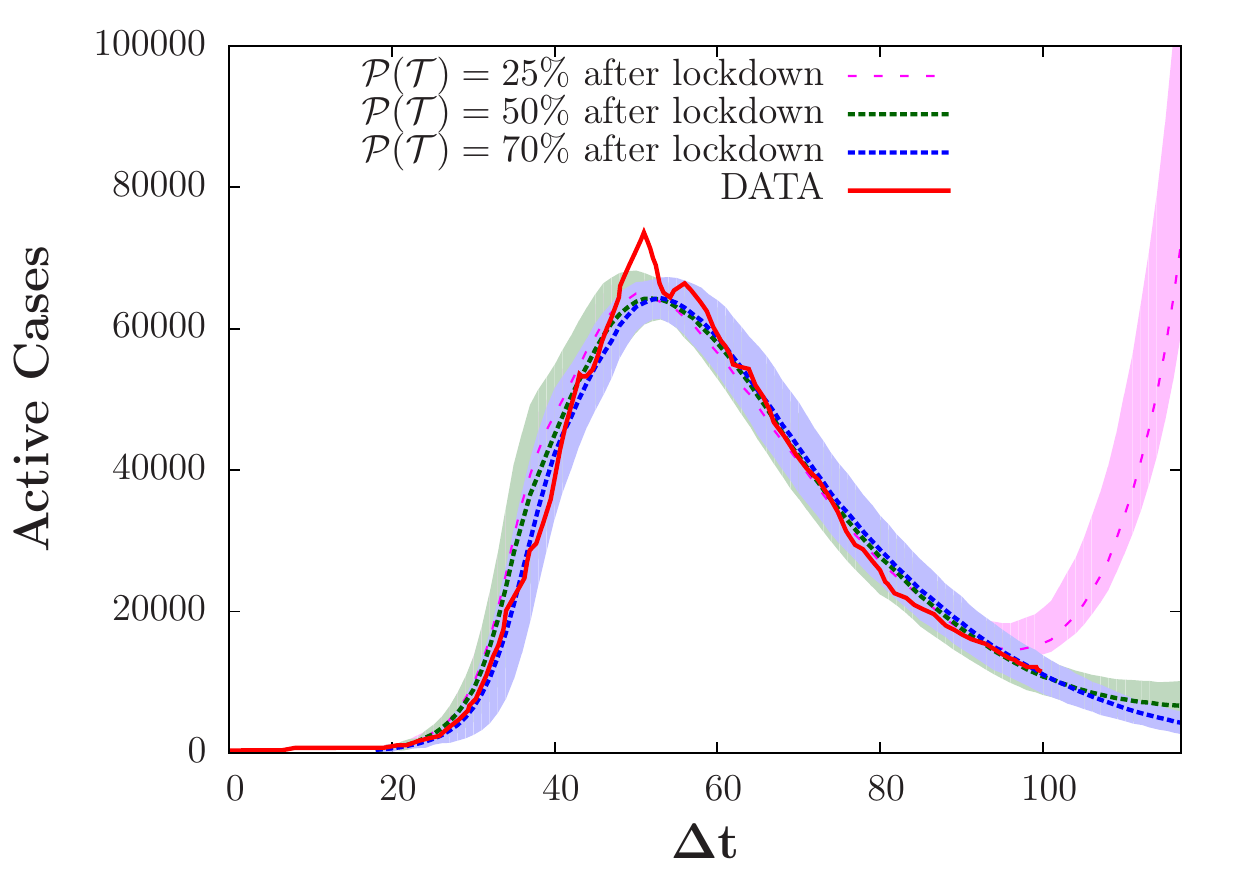}
% ACGER.pdf: 0x0 px, 0dpi, nanxnan cm, bb=
\caption{Germany active cases }
\label{fig:ACGER}
    \end{subfigure}%
        \begin{subfigure}[b]{0.5\textwidth}
 \includegraphics[width=\textwidth]{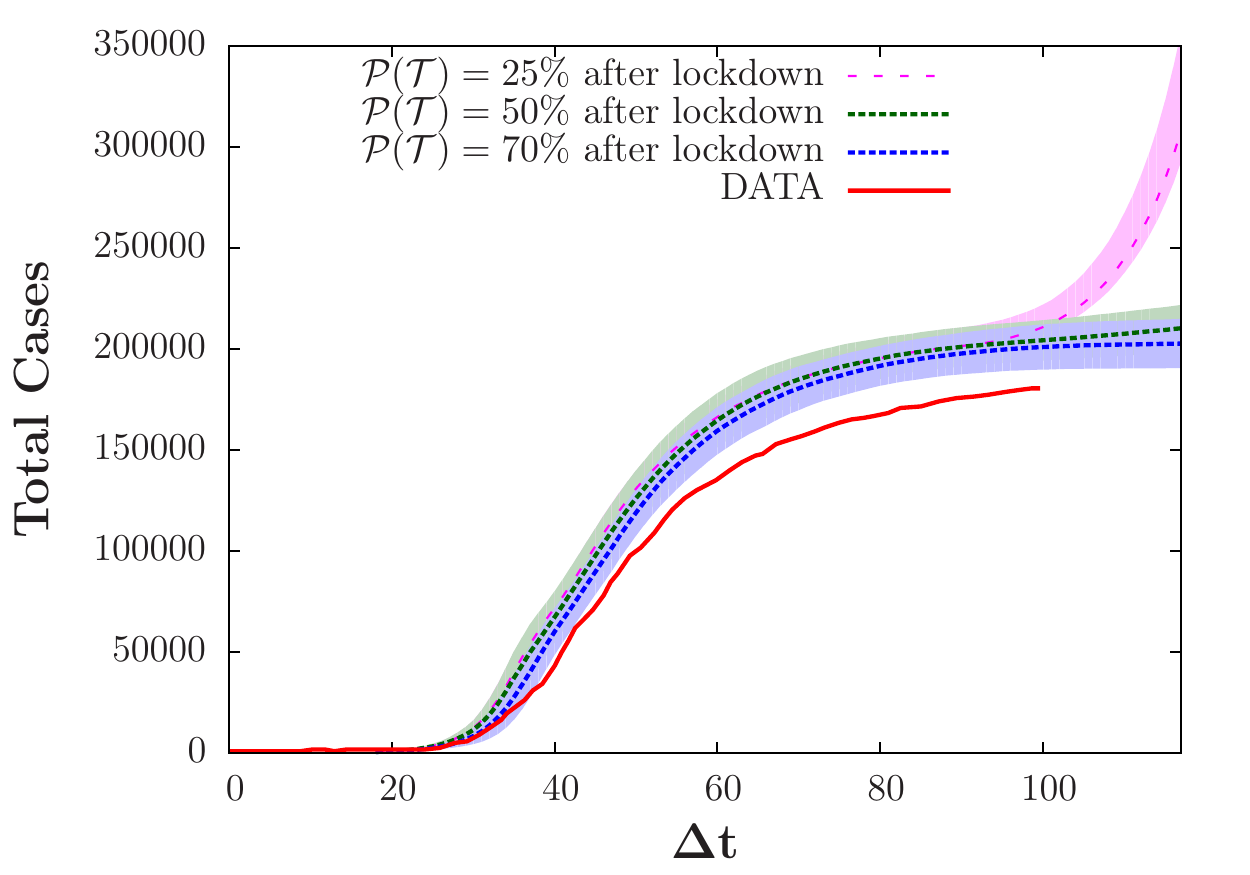}
% TCGER.pdf: 0x0 px, 0dpi, nanxnan cm, bb=
    \caption{Germany total cases }  
    \label{fig:TCGER}
    \end{subfigure}
    \caption{Comparison plot of Germany COVID-19 epidemic data with our simulation results obtained from SIR model over a random graph. The data compared here is from {\em $15th$ February 2020}. The random graph used have one million nodes with $40$ mean edges and averaged over $50$ different realizations. The parameters $\alpha=0.4$, $\gamma=1/14$ are used. The shaded regions represent the standard deviation with the mean values denoted by lines.} 
    \label{fig:GERMANY}
\end{figure}

Germany adopted different strategies compared to South Korea for controlling the spread. Where South Korea control relied on effective testing, Germany had a serious lockdown in the initial stages, not allowing even two people to stay together in the public spaces. In the later stage Germany adopted testing and contact tracing with high effectiveness, before opening the state by relaxing the lockdown measures. According to the available news sources, the serious lockdown was implemented on March, $22nd$, when the number of infectious was approximately $24,500$. In the simulation, lockdown with $85\%$ probability was implemented for a period of $55$ days, starting from day 36. After a week, testing with the effectiveness of $25\%$ was also introduced, when the infectious population was approximately $60000$; at this time a small change in the dynamics can be observed in the active and the total cases. Note that the testing effectiveness of the real system may be slightly different due to presence of asymptomatic infectious which are not fully captured in the real data. Before opening, we have employed testing with varying effectiveness, as shown in Fig.~\ref{fig:GERMANY}. If the system continue with testing with effectiveness of $25\%$, the model predicts a resurgence of second peak. Thus more efficient testing is needed for controlling the resurgence of the disease as can be seen in Fig.~\ref{fig:ACGER}. We find that the effectiveness of $50\%$ represents the real data well. The number of active cases obtained from the simulation follows the real data remarkably well. We also get a good agreement for the number of daily cases, as one can see in Fig.~\ref{fig:DCGER}. However, the total cases is higher in our simulation and can be attributed to the limitations inherent in the model for representing the real system, as well as a lack of sufficient information for implementing the mitigation strategies. Some discrepancies may arise due to under representation of the asymptomatic infectious in the real data.

%The active cases predicted by the model is in good agreement with the real data; however, the total cases is slightly overpredicted. This suggests higher rate of infection as well as recovery in the model as compared to the real system. The discrepancy may arise due to various limitations inherent in the model and a lack of sufficient information for implementing the mitigation strategies. We also note that in the real system, there are a large number of infected who are asympomatic and are not included in the data.  

\subsection{New York}
\label{NewYork}

\begin{figure}[!ht]
    \begin{subfigure}[b]{0.5\textwidth}            
    \includegraphics[width=\textwidth]{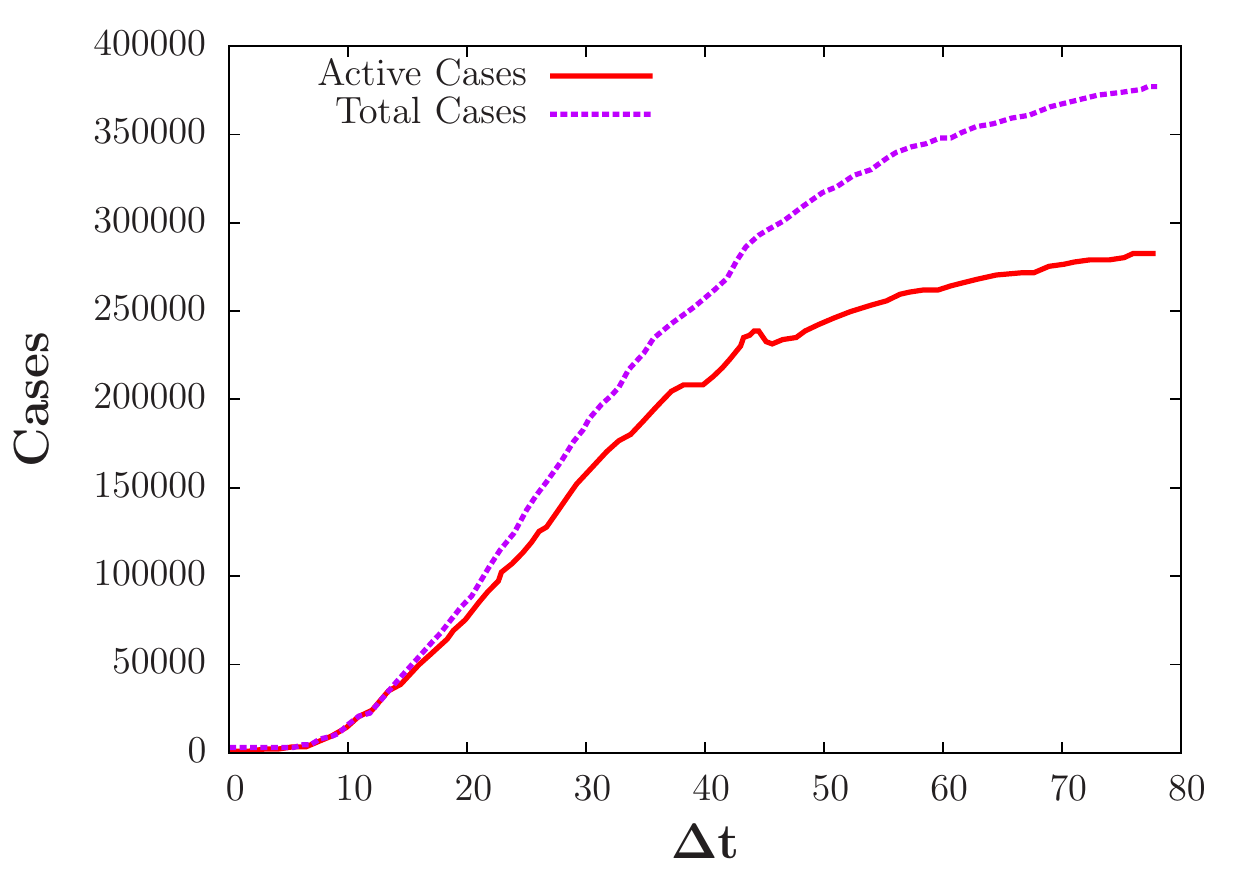}
% GD.pdf: 0x0 px, 0dpi, 0.00x0.00 cm, bb=
    \caption{New York COVID-19 data }
   \label{NY_data}
\end{subfigure}
\begin{subfigure}[b]{0.5\textwidth}            
    \includegraphics[width=\textwidth]{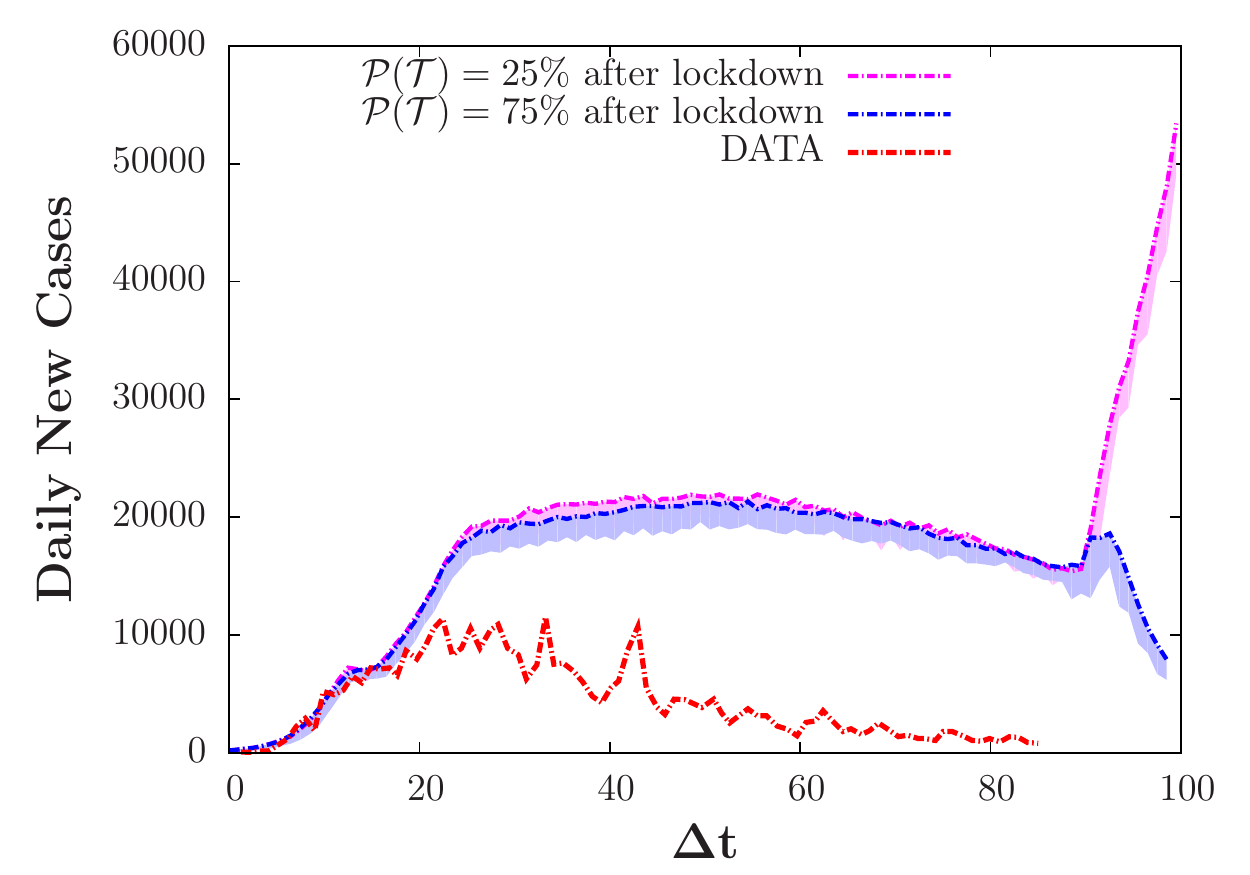}
     \caption{New York daily infected case}
    \label{DCNY}
\end{subfigure}
\begin{subfigure}[b]{0.5\textwidth}            
    \includegraphics[width=\textwidth]{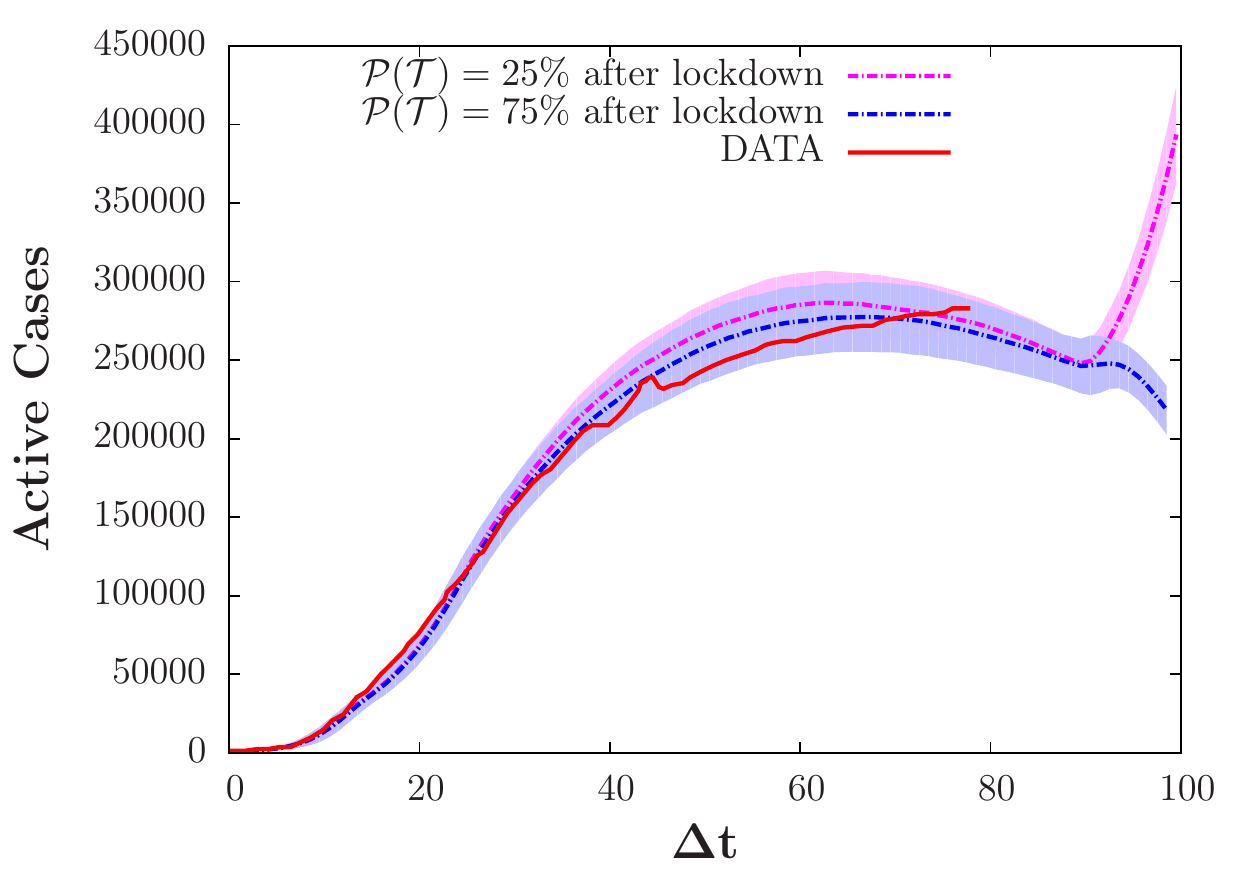}
% ACGER.pdf: 0x0 px, 0dpi, nanxnan cm, bb=
\caption{New York active cases }
\label{ACNY}
    \end{subfigure}%
        \begin{subfigure}[b]{0.5\textwidth}
 \includegraphics[width=\textwidth]{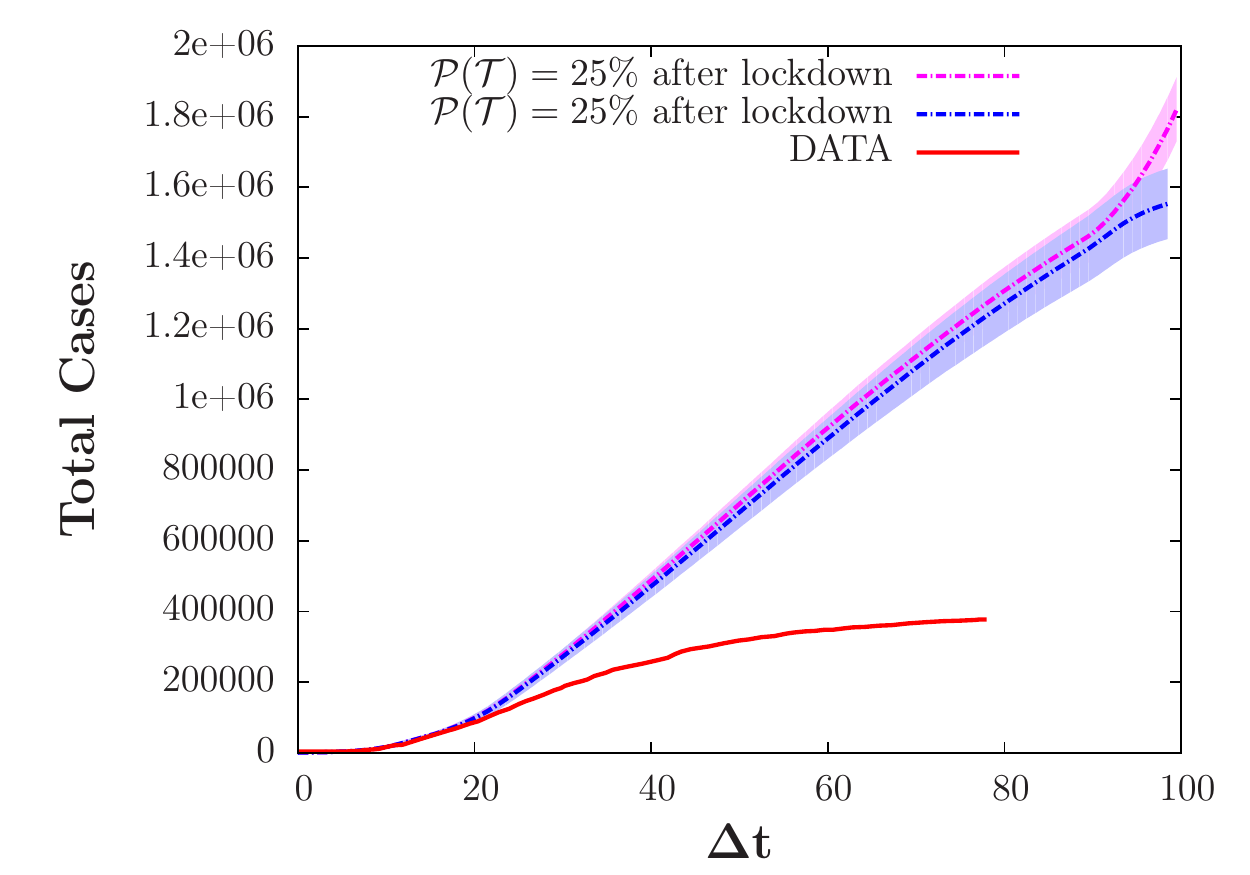}
% TCGER.pdf: 0x0 px, 0dpi, nanxnan cm, bb=
    \caption{New York total cases}  
    \label{TCNY}
    \end{subfigure}
    \caption{Comparison plot of New York COVID-19 epidemic data with our simulation results obtained from SIR model over a random graph. The data compared here is from {\em $12th$ March 2020}. The random graph used have one million nodes with $40$ mean edges and averaged over $10$ different realizations. The other parameters $\alpha=0.4$, $\gamma=1/14$ are used. The shaded region represent the standard deviation with its mean value denoted by lines.} 
    \label{fig:Newyork}
\end{figure}

New York is one of the highly populated states in the  United States of America, with a total population of approx $19.4$ millions. Due to being a densely populated and highly connected with the outside world, New York has the highest number of COVID-19 cases. It is one of the most challenging data to reproduce due to the limited information we can get from  IHME website \cite{COVID19NY}. 
According to IHME,  New York restricted all big gatherings by $18th$ of March and stayed at home order was passed on $22^{nd}$ of March. By $22^{nd}$ of March, the count of the active cases in New York was approximately $15,000$. Further, we do not have any proper knowledge when the testing was implemented in the system and what is its efficiency.
Similar to the other two countries, we start with choosing an infection rate such that the first ten days of available data matched for fixed recovery rate, $\gamma=1/14$ days$^{-1}$. In our model, the unmitigated infection rate of New York, $\alpha=0.4$, turns out to be the same as for South Korea with $10$ initial infections.  We then implemented the partial lockdown with a probability of $63\%$ when the number of infections is $15,000$. Later, in our model, we assumed that the testing was started by early April when the active cases are approximately $100,000$ and we implemented testing with the effectiveness of $25\%$ to capture the real data curves. We have implemented lockdown for $80$ days and after opening we study two cases; one where testing continue with $25\%$, and another the increase to $75\%$. Our study suggest that opening with efficient testing facilities will control the second outbreak as shown in Fig.~\ref{ACNY}. It is also important to note that the two different strategies \ref{SouthKorea} and \ref{NewYork} of the same system lead to very different growth behavior. We observe discrepancies in case total cases Fig.~\ref{TCNY} as well as in daily cases Fig.~\ref{DCNY}, results. These differences may be due to the Poisson distributed connections of the random graph, which limits the spread of the infections to such immense value with strong mitigation measures. We also observe that the daily cases number in simulation Fig.~\ref{DCNY} are almost double and continue to be large values for an extended period. It might be due to differences in the fixed recovery period we have chosen.

\section{Discussions and Conclusions}
\label{Sec:Concl}
In this study, a network model is utilized for studying the dynamics of infection spreading through a population. Here, we have used a random graph that solves the SIR model using parameters relevant for COVID-19. Furthermore, the SIR random network model has been extended to include the effect of mitigation strategies lockdown and testing. The network model including lockdown/test measure is then used to investigate the dynamics of epidemic growth in the mitigated system. In addition, using the developed models, we carry out three studies concerning the spread of infection in real systems (South Korea, Germany and New York) which utilize lockdown and testing measures for controlling the spread. We show that the network model including lockdown and testing measure preforms reasonably well in predicting the spread of COVID-19 in the real system studies. 

The lockdown measure is implemented by forcing the nodes to get isolated with a pre-specified probability; with probability one, the system is considered to be complete lockdown where the randomly picked nodes definitely isolated themself. The study of epidemic growth in a system under lockdown reveals a large peak in the number of infectious, similar to the uncontrolled dynamics, if the probability of the lockdown implementation is low (e.g. $25\%$). For implementations with large probabilities (e.g., $75\%$ or larger), the number of infected at the peak can become as large as that of an uncontrolled system if the lockdown is lifted and there is a second outbreak. The lockdown implementation with $50\%$ presents an optimal scenario in which there is a second outbreak and two peaks of equal amplitudes develop. However, at about $10\%$ of the population, the number of infected at the peaks can still be quite large. We also show that there is a second outbreak after the lockdown is withdrawn, which depends on the number of infected at the end of the lockdown period. If the number of susceptible is considerably larger than the herd immunity value then the system remain at the risk of the second outbreak, especially when the system can be reinfected due to outside interaction. 

The testing measure relies on the ability to identify the infectious as well as the asymptomatic in the population. The implementation depends on the finding of the contagious nodes with the effectiveness of testing and completely isolating them. We find that testing with effectiveness greater than $50\%$ works reasonable well for controlling the spread, and the peak in the infectious curve forms only at a slightly higher value than the level at which the measure is introduced. Beyond the peak value, the infectious curve starts dropping and if the mitigation is maintained long enough, the infection can be completely eradicated from the system. However, as we observed in case of lockdown, the system under testing remains at the risk of the second outbreak if a large number of susceptible exist in the population, especially when there is an outside interaction and any further testing is stopped.

Although both lockdown and testing rely on modifying the interaction between two individuals, it is essential to understand the underlying differences. For implementing the lockdown, no prior knowledge of the infected is required and all individuals are isolated, regardless of their status -- susceptible, infectious or recovered. Thus, the mitigation measure can be implemented easily and quickly without deploying large resources. For testing to be effective, one needs to establish facilities that can identify enough number of infected among a pool of potential spreader of the infection. Therefore, considerable effort and resources are required for implementing testing.      

Furthermore, testing is likely to perform better than lockdown in terms of compliance from the members of the population. In lockdown, all individuals are treated exactly the same, and regardless of their status as infectious, susceptible or recovered are kept in isolation. For this reason, the mitigation  measure is likely to suffers from the lack of compliance by the members of the population, especially when the implementation drags for a long time period. Testing, on the other hand, identifies and isolates the infected and those who are likely to catch the infection. Thus, only the potential spreads are affected, which makes the population's compliance possible to a large extent. Also, with testing the adverse effect on the economy and social lives of the people can be managed better.

Finally, we note that eliminating an infection as contagious as COVID-19 is difficult, primarily because the world populations are highly connected. Any remnant of the infections anywhere in the world has the potential to reignite the spread if the population has not acquire the herd immunity. Therefore, the goal is to reach the herd immunity, either through a vaccine or through the spread, preferably controlled. In the latter case, one needs to control the spread to a level that the increase in the infectious can be handled effectively by the local health-care facilities; at the same time, the number of the recovered is allowed to increase, ultimately attaining the herd immunity. Hence, the mixed strategies including both lockdown and testing will be useful in near-future. Various countries have used this approach; however, they differ in the many aspects such as the time to initiate the mitigations, lockdown period, the effectiveness of testing, etc. We have made the numerical comparison of a mixed mitigation scheme on the three very different state, all with very different mitigation strategies. Interestingly, with a very simplified model, we are able to reproduce the growth of the epidemic of these countries. Our study suggest that the resurgence in South Korea and Germany is controlled due to high efficient testing after opening and so New York should follow the same.  At this time, most of the countries are opening, and there might be a resurgence of the epidemic (also seen in Germany simulation when we decrease the testing efficiency). These comparative studies over real data give us a rough estimate of the combined effect of lockdown and testing. Thus similar studies using realistic models will be useful to gain insights of the epidemic spread.

\section*{Acknowledgments}
We would like to thank Uwe C. T{\"a}uber and Stefano Dell'Oro for valuable feedback. Priyanka Acknowledges the U.S. Army Research Office for financial support under Grant Number W911NF-17-1-0156. The views and conclusions contained in this document are those of the authors and should not be interpreted as representing the official policies, either expressed or implied, of the Army Research Office or the U.S. Government. The U.S. Government is authorized to reproduce and distribute reprints for Government purposes notwithstanding any copyright notation herein.
\newpage
 %\bibliography{SIR_epi}
%\bibliographystyle{unsrt}

\end{document}